\documentclass[sigconf,screen]{acmart}

\copyrightyear{2025}
\acmYear{2025}
\setcopyright{cc}
\acmConference[UIST '25]{The 38th Annual ACM Symposium on User Interface Software and Technology}{September 28-October 1, 2025}{Busan, Republic of Korea}
\acmBooktitle{The 38th Annual ACM Symposium on User Interface Software and Technology (UIST '25), September 28-October 1, 2025, Busan, Republic of Korea}\acmDOI{10.1145/3746059.3747789}
\acmISBN{979-8-4007-2037-6/2025/09}

\usepackage{xcolor}
\definecolor{MyLightGray}{HTML}{F2F2F2}

\newboolean{annotation}

\newcommand{\etal}{{et al.}}
\newcommand{\eg}{\textit{e.g.}}
\newcommand{\ie}{\textit{i.e.}}

\definecolor{red}{rgb}{0.8, 0.1, 0.1}
\definecolor{yellow}{rgb}{0.8, 0.6, 0.0}
\definecolor{blue}{rgb}{0.1, 0.2, 0.8}
\definecolor{pink}{rgb}{0.8, 0.0, 0.6}

\definecolor{BLUE}{rgb}{0.1, 0.2, 0.8}

\ifthenelse{\boolean{annotation}}{%
    
    \newcommand{\del}[1]{\textcolor{red}{[delete: #1]}}
    \newcommand{\com}[1]{\textcolor{yellow}{[comment: #1]}}
}{%
    
    \newcommand{\del}[1]{\ignorespaces}
    \newcommand{\com}[1]{\ignorespaces}
}

\input{floats.tex}

\begin{document}

\title{Cooperative Design Optimization through Natural Language Interaction}

\author{Ryogo Niwa}
\affiliation{
  \institution{OMRON SINIC X Corporation}
  \city{Tokyo}
  \country{Japan}
}
\affiliation{
  \institution{The University of Tsukuba}
  \city{Tsukuba}
  \country{Japan}
}
\email{niwa.ryogo.ms@alumni.tsukuba.ac.jp}
\orcid{0000-0003-0403-8738}

\author{Shigeo Yoshida}
\affiliation{
  \institution{OMRON SINIC X Corporation}
  \city{Tokyo}
  \country{Japan}
}
\email{shigeo.yoshida@sinicx.com}
\orcid{0000-0001-5381-0187}

\author{Yuki Koyama}
\affiliation{
  \institution{National Institute of Advanced Industrial Science and Technology (AIST)}
  \city{Tsukuba}
  \country{Japan}
}
\affiliation{
  \institution{The University of Tokyo}
  \city{Tokyo}
  \country{Japan}
}
\email{koyama@pe.t.u-tokyo.ac.jp}
\orcid{0000-0002-3978-1444}

\author{Yoshitaka Ushiku}
\affiliation{
  \institution{OMRON SINIC X Corporation}
  \city{Tokyo}
  \country{Japan}
}
\email{yoshitaka.ushiku@sinicx.com}
\orcid{0000-0002-9014-1389}

\begin{abstract}
  % !TEX root = ../paper.tex

Designing successful interactions requires identifying optimal design parameters.
To do so, designers often conduct iterative user testing and exploratory trial-and-error.
This involves balancing multiple objectives in a high-dimensional space, making the process time-consuming and cognitively demanding.
System-led optimization methods, such as those based on Bayesian optimization, can determine for designers which parameters to test next. 
However, they offer limited opportunities for designers to intervene in the optimization process, negatively impacting the designer’s experience.
We propose a design optimization framework that enables natural language interactions between designers and the optimization system, facilitating \emph{cooperative} design optimization.
This is achieved by integrating system-led optimization methods with Large Language Models (LLMs), allowing designers to intervene in the optimization process and better understand the system's reasoning.
Experimental results show that our method provides higher user agency than a system-led method and shows promising optimization performance compared to manual design.
It also matches the performance of an existing cooperative method with lower cognitive load.
\end{abstract}




\figteaser

\maketitle

% !TEX root = ../paper.tex

\section{Introduction}
\label{sec:introduction}

\paragraph{Background}
Designing successful interactions requires designers to carefully explore and identify optimal design parameters through iterative user testing.
Manually performing this task is challenging due to high-dimensional parameter spaces, where designers must consider both the relationships between design parameters and their impact on the resulting design performance.
Furthermore, there is a need to balance multiple objectives (\eg, speed, accuracy) simultaneously, as has been pointed out by researchers in human-computer interaction (HCI) \cite{Chan:CHI:22,Liao:PC:23,Bi:CHI:14,Yoshida:Access:24}.
Consequently, many iterations of user testing are required, demanding substantial time and cognitive effort, with no guarantee of finding the optimal solution.
To address these issues, researchers have developed human-in-the-loop design optimization approaches \cite{Chan:CHI:22,Liao:PC:23}, such as using \emph{Bayesian optimization} (BO) \cite{Shahriari:ProcIEEE:16} as a backend.
In these approaches, the system leads the optimization process---the system decides the parameters to test next with computational reasoning---providing a more principled and efficient procedure than those with manual decisions.

\paragraph{Motivation}
However, fully system-led optimization has been criticized for negatively impacting the design experience \cite{Chan:CHI:22}.
While designers often have their own ideas and intuitions about the exploration process, most system-led methods offer limited opportunities for them to intervene in the optimization process.
Without the ability to steer the optimization process, designers feel less sense of agency and also miss opportunities to gain deeper insights into the design problem.

\paragraph{Our Idea}
This leads to the question: can we foster more \emph{cooperative} interactions between designers and optimization systems by enabling communication through natural language?
Natural language offers the flexibility to steer the optimization process---even when designers have only a rough understanding of the design space.
Such an interface would allow designers to express their intentions in natural language and understand the system's reasoning behind its suggestions, thereby guiding the optimization process more effectively---capabilities that are often limited in standard BO, where the system leads the process with minimal room for intervention.
This design aligns with the principles of \emph{cooperative AI} \cite{Dafoe:Nature:21}, which emphasize shared understanding, communication, and commitment in human-AI collaboration.

\paragraph{Our Approach}
In this work, we propose the concept of \emph{cooperative design optimization through natural language} (\autoref{sec:interaction}).
To realize this, we present a novel technique to integrate an LLM with the BO-based optimization procedure (see \autoref{fig:teaser} and \autoref{sec:technique}), enabling designers to intervene through natural language and understand the system's intentions behind its suggestions.
We implement this technique in a working prototype system for designing web apps.
Through our user study with this prototype system (\autoref{sec:study:1}), we explore the potential of this approach to enhance the design experience compared to designer-led (manual) and system-led approaches.
Note that \citet{Mo:TIIS:24} recently proposed a cooperative design optimization system based on a custom BO technique, which allows designers to specify explicit constraints on parameter ranges to consider.
While we share the overall concept of cooperative design optimization, our approach differs in its interaction and implementation.
We incorporate an LLM into the BO process to support more flexible and interpretable interactions through natural language.
To understand the experiential differences between the two approaches---ours and that of Mo \etal---we conducted an additional user study (\autoref{sec:study:2}) and report the results.

\paragraph{Contributions}
The contributions of this work are threefold.
\begin{enumerate}
    \item We introduce the concept of cooperative design optimization through natural language, enabling flexible and interpretable interactions with the optimization process.
    \item We present a novel technique that integrates an LLM with the BO process, allowing the optimization behavior to be guided by natural language.
    \item We discuss the potential benefit of natural language interaction based on the results of the user studies, highlighting future challenges for improving the user experience.
\end{enumerate}

% !TEX root = ../paper.tex

\section{Related Work}

\subsection{Human-in-the-Loop Design Optimization}

In human-computer interaction (HCI), design optimization has typically been applied to two main problem settings.
The first focuses on maximizing subjective preferences, either from designers \cite{Koyama:SIGGRAPH:20,Yamamoto:UIST:22} or the general audience \cite{Koyama:SIGGRAPH:17}, often using preferential Bayesian optimization (PBO) as a key technique.
The second setting seeks to find the optimal interaction by maximizing user performance metrics, typically through iterative user testing.
In this context, \emph{adaptive experimental design} \cite{Yoshida:Access:24} is employed to reduce participant burden by adaptively adjusting testing conditions, rather than predefining them, thus minimizing the number of participants required.
BO is commonly used as the backend algorithm for adaptive experimental design \cite{Khajah:CHI:16,Yoshida:Access:24,Liao:PC:23,Dudley:CHI:19}.

Our work addresses the second setting: design optimization with iterative user testing.
However, fully system-led optimization, while computationally efficient, often overlooks the designer's intuition and experience, leading to a suboptimal designer experience \cite{Chan:CHI:22}.
To address this, various efforts have sought to give designers more control \cite{Koyama:UIST:22,Mo:TIIS:24}.
Our work builds on these efforts, with Mo \etal's cooperative system \cite{Mo:TIIS:24} being the most closely related.
However, our approach differs by incorporating LLMs to enable more flexible and interpretable natural language interactions, offering a novel approach to cooperative design optimization.

\subsection{Large Language Models for Optimization}

Large Language Models (LLMs) are highly effective at grasping context and utilizing their pre-existing knowledge, which makes them suitable for tackling optimization problems \cite{Huang:SEC:2024}. Some approaches delegate the entire optimization process to an LLM instead of using traditional algorithms. For instance, OPRO (Optimization by PROmpting) generates solutions solely through natural language prompts, removing the need for specialized optimization procedures \cite{yang2024large}. Another line of work describes the BO process in natural language so that an LLM can replicate the BO search steps \cite{Liu:ICLR:24}.

In addition to these fully LLM-based methods, hybrid strategies that embed an LLM into an existing optimization pipeline are gaining attention. While some employ frameworks other than BO \cite{Hao:SEC:2024}, many focus on integrating LLMs into BO. Our work follows this hybrid direction. For example, in one study, natural language descriptions of chemical experiments are transformed into a semantic feature space via an LLM and then used as input to BO for improving search performance \cite{rankovic:NeurIPS:2023}. Another study suggests replacing the standard surrogate model in molecular design tasks with an LLM, opening up new avenues for data-driven exploration \cite{Ramos:arXiv:23}.

More recently, several methods have combined an LLM directly into the BO sampling process. SLLMBO randomly switches between a Tree-structured Parzen Estimator (TPE) and an LLM-based sampling at each iteration, balancing exploration and exploitation \cite{Kanan:arXiv:2024}. Another method, BORA, dynamically selects from three sampling strategies based on the current context. Although these methods perform well on single-objective tasks, their effectiveness in multi-objective optimization remains unverified \cite{cisse:arXiv:2025}, and most of this work focuses primarily on algorithmic performance. In contrast, our approach highlights \emph{cooperative design optimization}, investigating how natural language interaction between designers and systems can enhance both user experience and optimization outcomes.

\subsection{Collaborative Optimization Modeling}

Recent studies view optimization as a collaborative process where humans and optimization systems interact to find solutions, and they investigate how to effectively support this interaction.
One research direction focuses on improving system transparency to enhance user trust. For instance, it has been suggested that, in addition to visualizing intermediate solutions, allowing users to directly verify outputs and engage in trial-and-error helps foster appropriate calibrated trust that reflects the system's actual behavior \cite{Liu:TIIS:2023}. While this prior work focuses on understanding and building designer trust in optimization systems, our research focuses on how using natural language for interaction with the optimization system affects the designer experience.

Other related studies use LLMs to bridge the communication gap between users and optimization systems. Some methods translate vague natural language inputs into formal constraints for optimization solvers \cite{Lawless:TIIS:2024,Michailidis:CP:2024,Li:arxiv:2023}. While these studies and ours both aim to enable natural language-based user input, they differ in terms of the optimization backend. Additionally, other approaches leverage LLMs to automate the optimization modeling process, which reduces the need for specialized expertise \cite{AhmadiTeshnizi:ICML:2024, Chenyu:arxiv:2025}. These approaches mainly address the formulation phase of optimization, whereas our approach intervenes during the execution phase by guiding the selection of candidate solutions.

% !TEX root = ../paper.tex

\section{Interaction: Cooperative Design Optimization through Natural Language Interaction}
\label{sec:interaction}

\subsection{Target Design Tasks}

The primary goal of this work is to identify optimal interactions by exploring and fine-tuning design parameters in the target interactive system.
We specifically focus on scenarios where this task is performed by UI/UX designers and HCI researchers refining their systems through iterative user testing \cite{Chan:CHI:22,Liao:CHI:24,Liao:PC:23,Mo:TIIS:24}.
Examples of such tasks include optimizing hand-based pointing interactions \cite{Chan:CHI:22,Liao:CHI:24}, configuring haptic notifications on smartwatches \cite{Liao:PC:23}, and adjusting the visual design of web pages \cite{Mo:TIIS:24}.
In these scenarios, designers typically manipulate $n$ continuous design parameters within finite value ranges, often controlled via sliders.
The objective is to maximize $m$ performance metrics, represented by objective functions $f_{1}, \ldots, f_{m}$.
Examples of such metrics include accuracy, speed, and subjective ratings based on aesthetic preference.

Each objective function is evaluated through user testing, where the performance metrics are measured.
Since user testing is labor-intensive and time-consuming, it is crucial to minimize the number of required tests.
This requires selecting informative test conditions (\ie, parameter configurations to be tested) for each iteration.
The ultimate goal is to identify a \emph{Pareto front}, a set of solutions where no objective can be improved without worsening another, allowing designers to understand the trade-offs between competing objectives.
The name ``front'' comes from how these solutions form a boundary when plotted, showing the edge of non-dominated solutions in the objective space.
Once the Pareto front is identified, the designer selects a single solution from it based on their understanding of these trade-offs.

\subsection{Design Optimization Approaches}

The simplest approach to design optimization is for the designer to manually adjust the design parameters, which we refer to as the \emph{designer-led} method.
This approach allows designers to freely explore the design space based on their intuition, while also deepening their understanding of the target design task.
However, it has been pointed out that this method makes it difficult to navigate high-dimensional design spaces, often leading to \emph{fixation}, where designers repeatedly explore the same areas of the design space without making significant progress \cite{Chan:CHI:22}.

In contrast, \emph{system-led} design optimization methods, such as those using BO, allow for more efficient and broad exploration of the design space \cite{Chan:CHI:22}.
However, these methods limit the designer's ability to intervene in the optimization process, which can result in a poorer designer experience, particularly in terms of the designer's sense of agency \cite{Chan:CHI:22}.

Our approach, \emph{cooperative} design optimization \cite{Mo:TIIS:24}, aims to create a flexible balance of control between the designer and the system.
Similar to system-led optimization, where the designer receives suggested parameter values from the system, conducts evaluation experiments, and measures performance, cooperative design optimization allows designers to intervene in the system's behavior and make final decisions.
This approach not only leverages the computational efficiency of system-led optimization but also enables designers to apply their knowledge and intuition throughout the process.

\subsection{Cooperation through Natural Language}

Previous work \cite{Mo:TIIS:24} introduced a specialized cooperative BO technique that allows designers to constrain the search space by explicitly specifying constraints (called ``forbidden regions'' and ``forbidden ranges'').
While this method enables the designer to control the optimization process, it requires careful consideration and precise input of specific values from the high-dimensional search space, which can be burdensome.

Our proposed cooperative interaction differs from this approach in two key ways:
\begin{enumerate}
    \item Designers can intervene in the sampling process (\ie, generating new parameters to test) using natural language. This allows them to provide general requests in natural language to guide the optimization, making the system-led optimization process more cooperative.
    \item The system provides natural language explanations for why the specific parameter values were proposed. This feature helps designers better understand the system's intention, enabling them to plan the next steps in the optimization process more effectively.
\end{enumerate}
Additionally, designers can manually adjust the parameter values to test next, either by selecting values entirely on their own or by modifying the system's suggestions.
As in the previous work, this flexibility allows designers to take full control of the optimization process when desired.

\subsection{System Interface and Interaction Scenarios}

To facilitate effective interaction with the LLM, our system offers the following features:
\begin{itemize}
    \item A text box for specifying requests (if any) to the LLM,
    \item A button to calculate parameter proposals based on the requests, and
    \item A display area for showing the LLM's reasoning.
\end{itemize}
In addition to these core features, our system inherits functionalities from the previous system \cite{Mo:TIIS:24}, enabling designers to interact with the optimization process directly:
\begin{itemize}
    \item Slider widget: This widget provides a set of sliders corresponding to the target parameters. Designers can use it to directly specify the set of parameters to test next. It also reflects the parameter values proposed by the system immediately after each suggestion, helping designers see exactly what values were proposed.
    \item Parallel coordinates plot (PCP): This plot visualizes the explored parameter sets, helping designers understand the relationships between parameters. The PCP is linked to the objective chart, allowing designers to easily see how well each parameter set performs.
    \item Objective chart: This chart visualizes the observed objective (performance) values of the tested parameter sets. It also highlights the most recently evaluated parameter set, making it easier for designers to track and confirm recent results.
\end{itemize}
Refer to \autoref{fig:user_interface} for a screenshot of the implementation used in our study (note that the system in the screenshot contains some user study-specific features such as \emph{formal} and \emph{informal} evaluation buttons; we will describe these later).

\autoref{fig:teaser} illustrates a typical interaction scenario of LLM-based cooperative design optimization.
For example, in a web design scenario, a designer may express a preference such as ``I want to explore variations with larger font sizes'' or provide an instruction based on performance trade-offs, like ``I want to find a design that prioritizes speed over accuracy.''
The system then generates parameter suggestions that reflect these requests.

Moreover, when the designer is uncertain about the rationale behind a suggested parameter set, they can read the reasoning by the system.
For instance, the system might provide an explanation: ``This suggestion focuses on maximizing accuracy in an area of the design space that has not yet been explored.''
This explanation helps the designer understand the system's intentions, allowing them to decide to either accept the proposed parameters for further testing, modify them based on their intuition, or provide a different request for generating alternative parameters.

% !TEX root = ../paper.tex

\section{Technique: LLM-Guided Bayesian Optimization}
\label{sec:technique}

\figsystemflow

We integrate an LLM into the BO process to realize cooperative design optimization through natural language interaction.
In each iteration, the system selects a single point (\ie, a set of parameter values) to be evaluated next with the help of an LLM (\autoref{sec:tech:llm}) from multiple candidates generated by a \emph{batch} BO technique (\autoref{sec:tech:bo}).
In this way, we can integrate the designer's request into the optimization process.
See \autoref{fig:system_flow} for an overview of the system procedure.

\subsection{Problem Formulation}

By iteratively conducting user testing, we want to solve the following multi-objective optimization problem:
\begin{align}
    \max_{\mathbf{x} \in \mathcal{X}} [ f_{1}(\mathbf{x}), \ldots, f_{m}(\mathbf{x}) ],
\end{align}
where $\mathbf{x}$ is a vector of $n$ design parameters to adjust, $\mathcal{X} \subset \mathbb{R}^{n}$ is the search space, and $f_{i} : \mathcal{X} \to \mathbb{R}$ is the $i$-th objective function (performance metrics such as speed and accuracy).
In each step, we choose a single parameter set $\mathbf{x}'$, conduct a user test using $\mathbf{x}'$, and then observe performance values $f_{1}(\mathbf{x}'), \ldots, f_{m}(\mathbf{x}')$.
The final outcome of the optimization process is a Pareto front.

\subsection{Batch Bayesian Optimization-Based Generation of Next Candidates}
\label{sec:tech:bo}

\paragraph{Surrogate Model}
Following the standard choice, we use a Gaussian Process (GP) as the surrogate model for each objective.
Our implementation uses the default setting of BoTorch \cite{Balandat:2020:botorch} for this purpose.

\paragraph{Acquisition Function}
An acquisition function is a measure that BO uses to decide which points to sample next. 
Although any acquisition function can be employed in principle, we use \texttt{qLogNEHVI} \cite{Daulton:NIPS:23} as our acquisition function, which is a state-of-the-art acquisition function for multi-objective problems.
It is an extension of \texttt{NEHVI} (Noisy Expected Hypervolume Improvement), which uses the expected improvement in hypervolume to locate promising candidate points while accounting for predictive uncertainty.
\texttt{qLogNEHVI} extends \texttt{NEHVI} to batch sampling (\ie, selecting multiple points that have high acquisition values and maintain diversity), so it can propose multiple candidate points at once in each iteration.

\paragraph{Batch Sampling}
Using the batch BO technique, we sample $q$ candidate points $\{ \textbf{x}^1, \textbf{x}^2, \ldots, \textbf{x}^q \}$ in each iteration.
In standard batch BO settings, these candidate points are expected to be evaluated in parallel.
However, our approach uses these candidate points differently;
we use the LLM to select only one point from the $q$ candidates and then conduct the evaluation of the selected point.
We repeat this process, feeding the observed objective values back into the GP and incorporating the designer's request at each step.

\subsection{LLM-Guided Candidate Selection and Reasoning}
\label{sec:tech:llm}

Among the $q$ candidate points generated through the batch BO technique, the LLM identifies the point that best aligns with the designer's natural language requests.
It also generates a natural language explanation of its decision so that the designer can understand why the system chose those particular parameter values.

\subsubsection{Prompt}
The prompt given to the LLM begins with the main instruction: 
``Based on the user's request described below, select the index of the candidate point and provide a reason for your choice.''
This is followed by the following information.
\begin{description}
    \item[Task Information] Explanation of the design task, including the overall description, the meaning and range of each parameter, and the meaning and range of each objective function.
    \item[Candidate Points] Each candidate point contains the acquisition function value, as well as the predictive mean and variance (potentially unnormalized) from the GP. When necessary, we revert these values to the same scale the designer sees in the interface. If the unnormalized predictive mean exceeds its defined range, we clip it to the maximum (or minimum) to avoid confusion for the designer.
    \item[Evaluation History] List of the previous observations. Each observation is a pair of the parameter set and its observed objective values.
    \item[Designer's Request] A sentence provided by the designer to express their preferences or goals. For example, they might write, ``I want the interface to be more visually accessible,'' or give a brief and abstract guideline. If left blank (\ie, the text box is empty), this part of the prompt is also left blank.
\end{description}
We provide more details in \autoref{sec:appendix:prompt}.
Furthermore, a technical evaluation of how faithfully the LLM selects candidates according to the designer's request is discussed in \autoref{sec:appendix:tech-eval}.

\subsubsection{LLM Outputs}
Given the prompt, the LLM outputs the index of the chosen candidate point and the reason for that choice.
For example, it might say: ``To meet the goal of increasing user revenue, I selected the candidate that uses a larger number of ads.''

\subsection{Overall Procedure}

In summary, the overall procedure is as follows.
\begin{enumerate}
\item \textbf{Obtain Initial Samples}\quad Begin the BO process with a small set of parameter sets (chosen randomly or via an experimental design).
\item \textbf{Batch Sampling}\quad Train a surrogate model for each objective function using the observed data, and then use the batch BO technique to sample $q$ candidate points $\{ \textbf{x}^1, \textbf{x}^2, \ldots, \textbf{x}^q \}$.
\item \textbf{Selection by the LLM}\quad Ask the LLM to select the single best point from the $q$ candidates that best aligns with the designer's requests and to generate an explanation of its decision in natural language.
\item \textbf{Evaluation and Update}\quad
Evaluate the selected point via user testing, observe the objective values, and then re-train the surrogate models with this new data.
\item \textbf{Iteration}\quad Repeat Steps 2--4.
\end{enumerate}

% !TEX root = ../paper.tex

\section{User Study 1: Comparing Levels of Control}
\label{sec:study:1}

This study aims to examine how differences in the levels of control affect both user experience and optimization performance in a design optimization task.
Participants were asked to complete design optimization tasks for web applications using three distinct ways: entirely relying on the designer (\textit{Designer-led}), entirely relying on the optimizer (\textit{BO-led}), and our proposed method, which enables cooperation with the optimization system through natural language (\textit{Cooperative (Natural Language)}).

We formed the following hypotheses:
\begin{itemize}
    \item \textbf{Hypothesis 1.1 (H1.1)}: Providing natural language instructions to the optimization system and receiving explanations for its suggestions improve users' sense of agency on design optimization tasks.
    \item \textbf{Hypothesis 1.2 (H1.2)}: Providing natural language instructions to the optimization system and receiving explanations for its suggestions improve optimization performance.
\end{itemize}

This study was approved by the ethics review board of the authors' institution.

\subsection{Study Design}

Our study design largely followed the approach used in previous work \cite{Mo:TIIS:24} by Mo \etal\, and we customized their publicly available system\footnote{\url{https://github.com/georgemo535/D3MOBO}} for our study.
The key difference here is that we used our proposed system for the cooperative optimization condition instead of theirs.

\subsubsection{Condition}

To ensure that each participant could experience all three approaches firsthand and to facilitate direct comparisons of agency, user experience, and optimization performance, we structured our study with a within-participant design. Specifically, within this design, each participant was exposed to three conditions: \textit{Designer-led}, \textit{BO-led}, and \textit{Cooperative: Natural Language}.
\begin{description}
    \item[Designer-led]
        Participants were able to adjust the parameters by manipulating sliders on the UI.
    \item[BO-led]
        Participants received parameter suggestions from a multi-objective BO system and were not allowed to adjust the values themselves using sliders.
    \item[Cooperative: Natural Language]
        Participants could both adjust parameters using sliders and request suggestions from the LLM.
        They were able to communicate with LLM through natural language, such as specifying the type of parameters they wanted or asking for clarification about the suggestions.
        In this section, we refer to this condition simply as \textit{Cooperative}.
\end{description}
The order of conditions and design tasks was counterbalanced across participants to mitigate potential order effects.

\subsubsection{Target Design Tasks and Simulated User Testing with Synthetic Test Functions}
\label{sec:target_design_tasks_and_simulated_user_testing}

We wanted participants to use interfaces under different conditions and reflect on their relative advantages and disadvantages.
To achieve this, it was necessary to conduct the study using a within-participant design.
Additionally, different tasks were required for each condition to prevent learning effects.
However, conducting \emph{actual} user testing in a design optimization task is time-consuming and makes it challenging to quantitatively compare the difficulty of each prepared optimization task.

To address this challenge, we adopted an alternative evaluation method following previous work \cite{Mo:TIIS:24}: instead of actually performing user testing, simulate it using \emph{synthetic} test functions.
We also adopted their design optimization tasks for three web applications (see \autoref{fig:app_screenshots} for screenshots).
This approach could save the time needed to perform each design task, making our within-participant study design feasible.

In all three web applications, the number of design parameters is five ($n = 5$), and the number of objective functions is two ($m = 2$).
The design parameters include, for example, ``icon transparency'', ``icon size'', ``text size'', and so on.
The objectives differ among the three web applications, but typical examples are ``average speed'' and ``accuracy''.
Refer to \autoref{sec:appendix:tasks} for the full details of parameters and objectives.
In the optimization process, parameters are normalized to $[0, 1]$, and objectives are normalized to $[-1, 1]$.
Thus, all objective functions have a domain of $[0, 1]^5$ and a range of $[-1, 1]^2$.

Mo \etal\ aimed to ensure that the difficulty in finding the optimal trade-off design across these test functions is nearly identical.
Therefore, they designed the functions to have similar forms with the same final Pareto hypervolume.
In realistic design scenarios, the objective functions are expected to be nearly convex, and multiple modes are rare. The behavior of the objective functions in prior research can justify this \cite{Dudley:CHI:19}.
Thus, they designed the test functions as quadratic functions for each objective ($j=1, 2$), where the optimal value is $c_j$, the position of the optimum is $\mathbf{a}_j=\left[a_{j1}, a_{j2}, a_{j3}, a_{j4}, a_{j5}\right]$, and the scaling factors are $\mathbf{b}_j=\left[b_{j1}, b_{j2}, b_{j3}, b_{j4}, b_{j5}\right]$.
Specifically, the function has the form:
\begin{equation}
\label{eq:evaluation_function}
f_j(\mathbf{x})=c_j-\sum_{i=1}^5 b_{j i}\left(x_i-a_{j i}\right)^2.
\end{equation}
Exact values of $\mathbf{a}_j, \mathbf{b}_j, c_j$ are provided in \autoref{sec:appendix:tasks}.

\figappscreenshots

\subsubsection{Formal and Informal Evaluations}\label{sec:user_study:study_design:formal_and_informal_evaluations}

Following previous work \cite{Mo:TIIS:24}, we prepared two types of evaluation methods: \emph{formal} and \emph{informal} evaluations.\footnote{\citet{Mo:TIIS:24} called \emph{heuristic} evaluation, rather than \emph{informal} evaluation; these terms indicate the same concept.}
Formal evaluation simulates rigorous user testing to measure target performance metrics accurately.
In contrast, informal evaluation simulates quicker but less accurate testing (\eg, conducted with fewer test users).
While the results of formal evaluations are used in the optimization process (in the BO-led and Cooperative conditions), the results of informal evaluations are not used for optimization;
instead, they help participants better understand the design space and plan which parameter set to evaluate formally next.

To simulate the timing and accuracy differences between these two evaluation methods, we applied the following artificial delay and noise:
\begin{itemize}
    \item For formal evaluations, we added small uniform noise sampled from $\operatorname{Uniform}(-0.05, 0.05)$ to the performance metrics. The result was made available 20 seconds after the participant requested a formal evaluation.
    \item For informal evaluations, we added larger uniform noise sampled from $\operatorname{Uniform}(-0.25, 0.25)$ to the performance metrics. The result was made available 3 seconds after the participant requested an informal evaluation.
\end{itemize}
We chose these specific noise ranges and delays, following \citet{Mo:TIIS:24}, to reflect typical differences in testing rigor and resource usage.

\subsubsection{Initial Seed}
As Mo \etal\ have noted, the system's behavior during the first few iterations can be unreliable due to limited data, and it is therefore advisable not to rely on the system at such an early stage. 
To address this, random sampling is often used during the first few iterations \cite{Frazier:arxiv:2018}.
Recent studies have specifically set the initial seeding to five iterations \cite{Liu:ICLR:24, Meinhardt:CHI:2025}. 
Our preliminary observations also suggested that the LLM's behavior stabilizes after about five iterations.
Based on these considerations, we decided to disable the parameter proposal function during the first five iterations in our implementation.

\subsubsection{Number of Samples per Batch}
A larger batch size allows our method to choose from a broader set of candidate points, which increases the chance of finding one that aligns with the designer’s request. However, setting it too high increases computational overhead and ultimately delays when the system can propose a new option that meets the designer’s needs. To maintain a balance between variety and responsiveness, we decided, based on empirical observations, to set the batch size to $q=8$.

\subsection{User Interface for Cooperative Condition through Natural Language}
\label{sec:user_study:study_design:user_interface}

\figuserinterface

\autoref{fig:user_interface} illustrates the interface used in the Cooperative condition, which consists of three panels and five components.

\subsubsection{Set Parameters Panel}

\autoref{fig:user_interface} (a) shows a set of sliders that allow participants to manually adjust design parameters at any time.
Each slider corresponds to a specific parameter, and changes are immediately reflected in the application.
For instance, in App 3 (Restaurant Map, \autoref{fig:app_screenshots} (c)), increasing the parameter ``Location Icon Size'' enlarges the red icon displayed at the center of the map.

\autoref{fig:user_interface} (b) includes a text box and an ``Ask AI for a New Design'' button. 
Participants can convey their preferences to the optimization system by entering natural language requests in the text box and pressing the ``Ask AI for a New Design'' button. 
Even without entering a request in the text box, pressing the button retrieves parameter suggestions from the system.
The suggested parameter values are then automatically applied to the slider.

The panel also provides a history of interactions with the optimization system in natural language.
Each time parameter values are retrieved, they are accompanied by an explanation describing the reasoning behind the suggestion.

\subsubsection{Evaluate Panel}

\autoref{fig:user_interface} (c) shows the buttons for informal and formal evaluation. 
Participants can evaluate the current parameter values on the slider by pressing either button.

\subsubsection{Check and Analyze Results Panel}

\autoref{fig:user_interface} (d) shows the objective chart, which visualizes the observed objective values obtained from formal and informal evaluations. 
Results obtained through informal evaluations are plotted in blue circles, while those from formal evaluations are shown as orange circles.
The most recent informal and formal evaluation results are highlighted using solid blue and orange dots.
The Pareto front is visualized by connecting the relevant points with green lines.
A checkbox allows users to toggle the visibility of the results from informal evaluations.

\autoref{fig:user_interface} (e) shows the parallel coordinates plot (PCP), which visualizes all previously explored parameter values.
Parameters evaluated through informal evaluation are visualized with blue dotted lines, while those evaluated through formal evaluation are shown as orange dotted lines. 
The most recent informal and informal evaluation parameters are visualized with solid blue and orange lines. 
Clicking on any of these dotted or solid lines restores the corresponding parameter values on the slider to help participants easily revisit previously explored parameters.

\subsection{Participants}
We recruited 18 participants (11 males and 7 females; average age: 28.94, standard deviation: 9.88).
All participants had prior experience related to UI/UX design: 11 had practical experience in UI/UX, 6 had experience in software development, and 1 had academic experience.
We specifically recruited participants with prior experience in UI/UX design through a social media post and personal outreach.
They were paid around \$17 for the compensation.
The user study was conducted remotely, with communication via Zoom, and participants performed the design tasks through a web browser.

\subsection{Procedure}
\label{sec:user_study_1:procedure}
The following outlines the steps of the procedure in the user study.
The entire study lasted approximately 1.5 to 2 hours.

\begin{enumerate}
    \item
    The study began with a guidance phase, where participants were shown a detailed video explaining the concept of parameter optimization and the idea of Pareto front.
    This was to familiarize the participants with the context of multi-objective design.
    We also told the participants about the context of designing a web interface for a technology company as a design optimization task.
    \item
    Following this, they watched a video tutorial explaining the ways they would engage in the design task to ensure the participants utilize the interactive features of the provided interface for the task. For example, if the task was under the Designer-led condition, the participants were introduced to the idea that an informal evaluation can provide an intuitive understanding of the optimization task. For the BO-led condition, they were introduced that parameters can be obtained by pressing the ``Get Parameters from System'' button. For the Cooperative condition, they were introduced to the feature that parameters can be obtained by pressing the ``Ask AI for a New Design'' button, and the reasons for the suggestions are also displayed.
    \item
    After viewing the tutorial video, participants completed a task that involved optimizing two design parameters based on two objective values.
    We gave the participants 5 minutes to engage in the tutorial task.
    During this phase, if participants did not interact with specific interface features, those features were explicitly explained to ensure a full understanding of the interface.
    \item
    Next, participants performed the main task, optimizing parameters in a web design application.
    They were instructed to find three Pareto-optimal solutions during this task.
    The task was capped at 20 minutes, but participants could finish early after 15 minutes.
    \item
    Upon completing each task, participants were asked to fill out a post-task questionnaire related to the specific condition they had just experienced.
    The questionnaire was designed based on the previous studies. Specifically, items addressing design space understanding and exploration experience were developed using \cite{Mo:TIIS:24} as a foundation, while the questionnaire items regarding agency were adapted with reference to \cite{Wang:IUI:24}. 
    In the Cooperative condition, participants were asked additional questions regarding the specific functionalities of the system.
    They then returned to step (2) to proceed with the next condition.
    \item After completing all three conditions, participants answered a final questionnaire comparing all the interfaces they used.
    \item Finally, we conducted a 15-minute semi-structured interview with each participant to gather additional qualitative feedback about their experiences with the system and suggestions for improvement.
\end{enumerate}

\subsection{Results}

\subsubsection{Pareto Set Discovery}
\figuserstudyfirstparetofronts
\figuserstudyfirstrelativehypervolume
\autoref{fig:user_study_first_pareto_fronts} shows the various Pareto fronts obtained in the different conditions across the three applications. We visually examined the Pareto fronts and found that, across all applications, the BO-led condition (blue) consistently appears in the upper-right region, indicating higher hypervolumes. In App 1 and App 3, the Cooperative condition (red) lies between the BO-led (blue) and the Designer-led (green). However, in App 2, the Designer-led (green) and Cooperative (red) fronts occupy roughly the same region. This result indicates that in App 1 and App 3, the Cooperative condition achieved better optimization performance than the Designer-led condition, whereas in App 2, both performed comparably.

We computed six metrics regarding Pareto sets, as described in \cite{Mo:TIIS:24}, to quantitatively compare the three conditions based on the obtained Pareto sets. 
In this section, we focus on the relative hypervolume, while the other metrics are presented in \autoref{sec:appendix_additional_metrics_from_user_study_1}. We choose this metric because it is widely used in multi-objective optimization, offering a comprehensive comparison with an ideal Pareto front.

\autoref{fig:user_study_first_relative_hypervolume} shows a boxplot of the relative hypervolume for each participant under each condition, normalized against the maximum hypervolume obtained in the corresponding application. A Friedman test revealed a significant effect of condition ($p=0.000$). Multiple comparisons with Bonferroni correction indicated significant differences between the Designer-led and BO-led conditions ($p.\text{adj}=0.000$), as well as between the BO-led and Cooperative conditions ($p.\text{adj}=0.001$). No significant difference was found between the Designer-led and Cooperative conditions ($p.\text{adj}=0.071$).

\subsubsection{Subjective Experience}
\label{sec:user_study_1:results:subjective_experience}

After each condition ended, participants answered the questions shown in \autoref{sec:appendix_questionnaire_and_detailed_results_from_user_study_1}. 
The questionnaire was divided into two categories: (1) agency and (2) participants' experience in understanding and exploring the design space. 
We first checked whether each question item in both categories satisfied normality. 
If normality was confirmed, we conducted a repeated-measures ANOVA followed by paired t-tests with Bonferroni corrections as a post-hoc analysis.
If that analysis indicated significant differences among the conditions, we then performed paired t-tests with Bonferroni corrections.
If normality was not confirmed, we conducted a Friedman test followed by pairwise Wilcoxon signed-rank tests with Bonferroni corrections.
 We present the results below.

\figuserstudyfirstagencycomposite

\paragraph{Agency.}

To evaluate participants' overall sense of agency, we computed an Agency Score by aggregating responses to the three 7-point Likert scale items (Q1, Q2, Q3) proposed by \citet{Wang:IUI:24}.
Specifically, we calculated the score as 
\begin{equation}
\label{eq:composite_score}
\text{Composite Score} = Q1 - Q2 - Q3
\end{equation},
so higher values indicate a stronger sense of agency.
Regarding this score, the Friedman test indicated a significant difference among the three conditions ($p=0.001$). 
Post-hoc comparisons revealed significant differences between the Designer-led and BO-led conditions ($p.\textrm{adj}=0.002$) and between the BO-led and Cooperative conditions ($p.\textrm{adj}=0.002$), but not between the Designer-led and Cooperative conditions ($p.\textrm{adj}=1.000$).
For a more detailed analysis of Q1, Q2, and Q3, see \autoref{sec:appendix_user_study_1_agency_related_questionnaire}.

\paragraph{Design Space Understanding and Exploration Experience.}
First, we checked each question for normality. 
We found that Q1, Q2, Q3, Q4, Q6, and Q8 did not meet the normality assumption, so we applied Friedman tests to those items; for Q5, which met normality, we used a repeated-measures ANOVA. 
As a result, Q1 ($p=0.046$), Q2 ($p=0.032$), Q3 ($p=0.000$), Q4 ($p=0.003$), Q6 ($p=0.003$), and Q8 ($p=0.001$) showed significant differences, while Q5 ($p=0.765$) and Q7 ($p=0.327$) did not.

Next, we ran pairwise comparisons only for questions with significant effects. 
For Q1, only BO-led vs. Cooperative was significant ($p.\textrm{adj}=0.048$); Designer-led vs. BO-led and Designer-led vs. Cooperative showed no significant difference. 
For Q2, a significant difference emerged between Designer-led and BO-led ($p.\textrm{adj}=0.048$), with no differences among the other pairs. 
In Q3, we found significant differences for Designer-led vs. BO-led ($p.\textrm{adj}=0.000$) and BO-led vs. Cooperative ($p.\textrm{adj}=0.001$), but not for Designer-led vs. Cooperative. 
Q4 showed significance only in BO-led vs. Cooperative ($p.\textrm{adj}=0.005$). Similarly, Q6 showed a significant difference only between BO-led and Cooperative ($p.\textrm{adj}=0.028$). 
Lastly, Q8 revealed significant differences for Designer-led vs. Cooperative ($p.\textrm{adj}=0.037$) and BO-led vs. Cooperative ($p.\textrm{adj}=0.002$), while Designer-led vs. BO-led was not significant.

\paragraph{Preference.} 
Twelve participants preferred the Cooperative condition, three preferred the designer-led condition, and three preferred the BO-led condition.

\subsection{Discussion}

We examined how our proposed system compared to other approaches regarding agency and optimization performance.

\subsubsection{Cooperative Interaction Increases Sense of Agency}
We found a significant difference in the Agency Score between the BO-led and Cooperative conditions.
This result suggests that the proposed system gives users a greater sense of agency than the BO-led condition.
Therefore, \textbf{H1.1.} is supported.

Participants described the experience of the cooperative condition as flexible and collaborative, with comments such as \textit{``I can explore it manually when I want, and rely on the AI when I prefer''} (P16) and \textit{``It feels like we're working together''} (P3). 
Another participant noted, \textit{``My intuition and the AI’s proposal fit together nicely''} (P3), highlighting the cooperative nature of the proposed method.

\subsubsection{User-Aligned Selection Would Limit Optimization Performance}
When comparing the BO-led and Cooperative conditions, we observed a significant difference in optimization performance, with the BO-led condition achieving a higher performance.
Meanwhile, the Cooperative condition showed a higher median than the Designer-led condition, although the difference was not statistically significant ($p.\textrm{adj} = 0.071$).
While these results indicate that \textbf{H1.2.} is not supported, these results suggest that the proposed cooperative system can achieve optimization performance comparable to or even exceeding manual human design.
Notably, the effect size for the comparison between the Cooperative and Designer-led conditions was large ($r=0.5333$), providing supportive evidence for the practical benefit of our cooperative approach.

From an algorithmic perspective, the BO-led condition sequentially selects the candidate with the highest acquisition function value in each iteration. 
In contrast, our proposed system uses batch Bayesian optimization to generate candidate points and then relies on the LLM to select a point that best aligns with the user’s natural language requests from the candidates. 
As a result, the system does not always choose the point with the highest acquisition function value, which may account for its lower optimization performance compared to the BO-led condition.

% !TEX root = ../paper.tex

\section{User Study 2: Comparing Cooperation Approaches}
\label{sec:study:2}

The aim of this experiment is to directly compare our approach with the existing work by Mo \etal \cite{Mo:TIIS:24}, which, like ours, pursues a cooperative optimization process but with a different cooperation approach.
We call their approach \emph{Explicit Constraint}: designers can intervene in the optimization process by explicitly specifying areas in the parameter space that they do not want to explore through a GUI.
Our proposed approach, by contrast, allows them to intervene more flexibly through natural language.

To evaluate the effectiveness of our approach in supporting human--AI collaboration, we focus on two key aspects: cognitive load and trust.
Natural language interaction enables designers to make ambiguous or flexible requests, eliminating the need to specify explicit constraints or fully understand the entire parameter space. 
This may help reduce cognitive load. 
In addition, providing explanations in natural language may help designers understand and accept the system's decisions, thereby fostering a greater sense of trust.

Based on these considerations, we formed the following hypotheses:
\begin{itemize}
    \item \textbf{Hypothesis 2.1 (H2.1)}: Our approach imposes a lower cognitive load than the Explicit Constraint condition.
    \item \textbf{Hypothesis 2.2 (H2.2)}: Our approach has higher perceived trustworthiness than the Explicit Constraint condition.
\end{itemize}

This study was approved by the ethics review board of the authors' institution.

\subsection{Study Design}
We largely followed the design and procedure from Study 1 but introduced several changes.
Specifically, the conditions differed, and we modified some of the questionnaire items.

\subsubsection{Condition}
We employed a within-participant design in Study 2, where each participant experienced two conditions.
\begin{description}
    \item[Cooperative: Explicit Constraint] Proposed by \citet{Mo:TIIS:24}, this method allows participants to obtain parameter suggestions either by adjusting sliders or through a customized BO system. Participants can also specify, via a GUI, which regions they do not want the system to explore.
    In this section, we refer to this condition as \textit{Cooperative-EC}.
    \item[Cooperative: Natural Language] This condition is the same as tested in Study 1.
    In this section, we refer to this condition as \textit{Cooperative-NL}.
\end{description}
We compared these two conditions across the three applications used in Study 1.
Each participant completed optimization tasks for two of the three applications, experiencing one condition per application.
The assignment of conditions to applications was counterbalanced across participants using a Latin Square design.

\subsubsection{Measurements}
To measure participants' trust in the system, we included items from the Multidimensional Trust Questionnaire (MTQ) \cite{Roesler:HFESAM2022}.
The MTQ assesses trust across four subdimensions---purpose, transparency, utility, and performance---using a 4-point Likert scale from ``Disagree'' to ``Agree.''
It captures trust in automated systems by targeting each subdimension with specific questions.
Because our task made it hard for participants to judge whether they achieved their goals, we omitted the performance dimension.
We also added the NASA-TLX \cite{Hart:NASATLX:1988} to measure the perceived workload of participants.

\subsection{Participants}
We recruited 12 participants (8 males and 4 females; average age: 25.4, standard deviation: 1.83).
Their backgrounds included 2 with practical experience in UI/UX, 8 with experience in software development, and 2 with academic experience.
We specifically recruited participants with experience in UI/UX design through a social media post and personal outreach.
They were paid around \$14 for the compensation. The user study was conducted remotely, with communication through Zoom, and the participants performed the design tasks through a web browser. No participants from Study 1 took part in this study.

\subsection{Result}

\subsubsection{Pareto Set Discovery}

\firuserstudysecondparetofronts

\autoref{fig:user_study_second_pareto_fronts} presents the Pareto fronts for the Cooperative-EC (blue) and Cooperative-NL (red) conditions across the three applications. Although there are minor differences among the applications, a visual inspection indicates that both conditions generally occupy similar regions in the Pareto space.

\firuserstudysecondrelativehypervolume

Following the methodology from Study 1 and building on Mo \etal\ \cite{Mo:TIIS:24}, we computed six metrics to quantitatively compare the resulting Pareto sets. 
Here, we focus on the relative hypervolume and refer readers to the appendix for the other metrics. 
\autoref{fig:user_study_second_relative_hypervolume} shows a boxplot of the relative hypervolume for both conditions. Because normality was not met, we conducted a Wilcoxon signed-rank test, which revealed no significant difference between the Cooperative-EC and Cooperative-NL) conditions ($p = 0.204$). 

\subsubsection{Subjective Experience}

Before analysing the data, we checked each measure for normality. We used a paired t-test for measures that satisfied the normality assumption and a Wilcoxon signed-rank test for those that did not. 

\firuserstudysecondnasatlxboxplot
\firuserstudysecondmtqboxplot

\paragraph{NASA-TLX}

We found a significantly lower weighted NASA-TLX score for the Cooperative-NL condition ($p=0.034$), as shown in \autoref{fig:user_study_second_nasa-tlx_boxplot}. Among the NASA-TLX subscales, we focused on mental demand (Q1) because it aligns most closely with our hypothesis on cognitive load. A statistical test on Q1 also revealed a significant difference ($p=0.000$), indicating lower mental demand under the Cooperative-NL condition.

\paragraph{Trust}
We asked participants to rate three subscales (Purpose, Transparency, and Utility), each consisting of three items on a 4-point Likert scale. We then averaged the scores within each subscale and applied Wilcoxon or paired t-tests, depending on normality. No significant differences emerged between the Cooperative-EC and Cooperative-NL conditions for Purpose ($p=0.090$), Transparency ($p=0.627$), and Utility ($p=0.347$).

\paragraph{Agency}
We used the same Agency Score as in Study 1, where the three 7-point Likert scale items (Q1, Q2, Q3) were aggregated according to \autoref{eq:composite_score}. Higher values indicate a stronger sense of agency. With only two conditions (Cooperative-EC and Cooperative-NL), we performed a paired t-test and found a significant difference ($p=0.032$) favoring Cooperative-EC. This result aligns with the observation that participants generally reported feeling more ``in charge'' under the Cooperative-EC condition (see \autoref{sec:appendix_user_study_2_agency_related_questionnaire} for a detailed analysis of Q1, Q2, and Q3).

\paragraph{Design Space Understanding and Exploration Experience}
We observed a significant difference in Q1 ($p=0.039$), favoring the Cooperative-EC condition, and in Q7 ($p=0.021$), favoring the Cooperative-NL condition. 
The remaining items, all tested with Wilcoxon (Q2: $p=0.142$, Q3: $p=0.151$, Q4: $p=0.431$, Q5: $p=0.131$, Q6: $p=0.469$, Q8: $p=0.394$), were not significant.

\paragraph{Preference}
Eight participants preferred the Cooperative-NL condition, while four participants preferred the Cooperative-EC condition.

\subsection{Discussion}
\label{sec:user_study_2:discussion}

\subsubsection{Reducing the Need for Parameter-Space Understanding Lowers Cognitive Load}
\label{sec:cognitive-load}

The NASA-TLX scores show that participants experienced significantly lower cognitive load with the Cooperative-NL condition than with the Cooperative-EC condition.
Therefore, \textbf{H2.1} is supported.

One participant (P2) said, \textit{``Even if I don't fully understand how each parameter affects the outcome, I could just say 'make it more accurate!''}, showing how focusing on the desired outcome can lower mental effort.

In contrast, the design interface in the Cooperative-EC condition would raise cognitive demands because it requires designers to specify forbidden regions, which multiple participants described as \textit{``cognitively demanding.''} 
One participant (P1) explained, \textit{``It felt like I had to do a lot of thinking. I needed to figure out `this parameter probably works better within that range' by myself, which took mental effort.''} 
Another participant (P10) added, \textit{``The interface (in the Cooperative-EC condition) was more mentally taxing. I had to think about which instructions to give, depending on the goal.''} 
Defining forbidden regions requires a certain level of parameter-space understanding, making it more burdensome in higher-dimensional problems.

Moreover, usage logs show that descriptions of desired outcomes were a prominent feature in the vast majority of requests (79 out of 85) in the Cooperative-NL condition. We provide a more detailed analysis of these queries in \autoref{sec:overall-discussion:analysis-of-requests}.
This focus on outcomes suggests that requesting the system \emph{``what''} to achieve, rather than \emph{``how''} to achieve it, would substantially reduce participants' cognitive load. 
Compared to specifying forbidden regions, merely requesting the desired outcome in natural language can lower designers' cognitive load.
While this goal-directed interaction is facilitated by the natural language interface, we acknowledge that this cognitive load reduction may come not only from the direct ease of natural language control but also partly from an automation effect, as designers are able to delegate the more complex task of parameter manipulation to the system.

\subsubsection{Why Natural Language Explanations Did Not Increase Trust as Expected}
\label{sec:system-trust}

We found no significant differences in average scores or individual questions on Transparency. The other two MTQ items (Purpose and Utility) also showed no significant differences in this study.
Therefore, \textbf{H2.2} is not supported.

We expected that providing natural language explanations of parameter suggestions in the Cooperative-NL approach would improve transparency and increase trust. 
As \autoref{fig:user_study_second_mtq_boxplot} shows, the medians for Trust, Utility, and Purpose all tended to be higher with Cooperative-NL, although the difference was not statistically significant. 

One possible reason is that some participants still felt uncertain about \emph{``how much their instructions were actually being followed,''} which may have prevented a stronger increase in trust. 
For instance, one participant (P7) commented, \textit{``We tried several things at first with the proposed method, but in the end, we only tweaked the AI’s suggestions. Meanwhile, with the second tool (Cooperative-EC), I could verify and adjust my requests all the time.''} Another participant (P12) said, \textit{``I trusted the second tool (Cooperative-EC) more overall, because I felt like it was really listening to me.''} 
These comments suggest that simply having the AI explain its reasoning is not enough,
making it difficult to be sure \emph{``the requests were really incorporated.''}

Cooperative-EC allows designers to visually confirm \emph{``which region is forbidden''} and see how the search proceeds, thus offering a concrete sense of control and reassurance. 
As one participant (P1) noted, \textit{``I can see the big picture, so I feel comfortable and I know what it's doing. That makes the first one I used (Cooperative-EC) feel more tangible to me.''} 
Another participant (P6) added, \textit{``After trying it out for a while, I get a sense of what works and can narrow things down by setting forbidden regions.''} 

While Cooperative-NL reduces cognitive load, it also provides fewer opportunities to deeply understand the parameter space---particularly in the early stages, when instructions may not be fully applied. 
As a result, some participants might have found it difficult to feel \emph{``in control,''} which may explain why Cooperative-EC scored higher on agency.
This factor likely contributed to the lack of a statistically significant difference in transparency, despite our assumption that natural language explanations would excel in this regard.

% !TEX root = ../paper.tex

\section{Overall Discussion}

This section reflects on the key findings from the two user studies, discusses their implications for cooperative design optimization.

\subsection{AI Suggestions Mitigate Design Fixation}
Design fixation \cite{Chan:CHI:22} is a phenomenon in which designers, while maintaining a strong sense of agency, can become overly attached to their initial ideas. In contrast, fully system-led approaches explore a wider range of possibilities but can undermine a designer's sense of agency.  
Our method addresses these challenges by providing unexpected yet feasible suggestions within the designer's chosen range. 
One participant commented:
\textit{``Even within the range I had explored, the AI suggested some surprising combinations that sometimes led to the best results. I had moments of 'Oh, I see!' that felt like real discoveries.''} (Study 1, P1).

Contradictory rationales from the system can prompt designers to question their assumptions, leading to a broader exploration of parameter settings. 
Another participant described how a conflicting rationale shifted their perspective:
\textit{``When I got a reason that differed from my expectations---like I wanted to increase user ratings, but it turned out to boost ads---I realized that even though I was trying to ignore revenue, the system was still increasing ads. Surprisingly, increasing ad parameters didn't hurt user ratings as much as I thought it would.''} (Study 2, P8).

These findings suggest that offering surprising yet valid suggestions can challenge habitual thinking and expand designers’ search space, helping mitigate design fixation while preserving a designer's agency.

\subsection{Explanations of AI's Reasoning Provide Acceptance and Reassurance}

Many participants noted that receiving ``reasons'' for parameter suggestions made it easier to plan subsequent parameter explorations. 
For example, one participant remarked,
\textit{``By having the system logically explain the causal relationship between parameters and outcomes, I could figure out which parameters to fix or adjust next. It was very convenient, and I felt it was better than my own approach.''} (Study 1, P9)
Similarly, another participant noted,
\textit{``There were times when it would explain the reasons for its suggestions each time, and based on that, I decided to give new requests.''} (Study 2, P9)

Some participants found explanations especially helpful when suggestions differed from their expectations, for instance:
\textit{``When I got a parameter suggestion different from my expectations, having an explanation made it easier to accept.''} (Study 2, P8). 
Others used these explanations to check that the AI's reasoning matched their own: 
\textit{``It reassured me that the AI's thinking aligns with mine.''} (Study 1, P4)

Taken together, these findings suggest that natural language explanations not only supported participants' decision-making during the design exploration but also helped them accept unexpected suggestions and feel reassured when the system's reasoning aligned with their own.

\subsection{Explanations for Parameter Suggestions Are Helpful but Need More Clarity and Depth}

While many participants found the natural language explanations helpful, some considered them unhelpful when they were overly general or too lengthy. 
For instance, \textit{``It just stayed with generic reasons''} (Study 1, P17) suggests that explanations lacking sufficient context or domain-specific insights may not offer meaningful guidance. 
Meanwhile, \textit{``I felt the explanation was too long''} (Study 1, P18) highlights the potential to increase cognitive load by our proposal.

One approach to address these issues is to structure the explanations in two layers: a concise summary, then an optional detailed view. 
Designers can quickly grasp the main points from the summary and refer to the more comprehensive explanation only if necessary. 
Furthermore, explicitly prompting the LLM to ``formulate hypotheses'' encourages more targeted and actionable reasoning, rather than just offering generic rationales.

\subsection{Designers Tend to Specify What, Not How}
\label{sec:overall-discussion:analysis-of-requests}

We analyzed the content of the 187 natural language requests from Study 1 and Study 2, manually applying multiple tags to each.
We used a two-category scheme: \textbf{Target}, to identify an instruction's focus (\texttt{Target: Outcome} or \texttt{Target: Parameter}), and \textbf{Type}, to describe its structural complexity.
The instruction types were \texttt{Type: Simple} (single-goal instructions), \texttt{Type: Compound} (instructions with multiple parallel goals), and \texttt{Type: Complex} (instructions with constraints or trade-offs difficult for GUIs to express).

This analysis revealed that designers primarily specified their goals (``what'') over the actions to achieve them (``how'').
Of the 187 requests, 170 (90.9\%) had a \texttt{Target: Outcome} tag, while only 31 had a \texttt{Target: Parameter} tag, with 26 containing both.
Furthermore, the instructions showed varied complexity.
Among the 170 requests classified as instructions, 66 (38.8\%) were \texttt{Type: Simple}, 69 (40.6\%) were \texttt{Type: Compound}, and 35 (20.6\%) were \texttt{Type: Complex}.
These \texttt{Type: Complex} requests included instructions that are difficult to express with sliders, such as weighting objectives (``ignore revenue and prioritize user ratings'') or imposing numerical constraints (``set the user rating to 4.0 or higher'').
The remaining 17 requests were categorized as \texttt{Type: Inquiry/Other}.
These results show that our natural language interaction could offer unique value, allowing designers to focus on high-level goals while flexibly articulating their complex intentions.

% !TEX root = ../paper.tex

\section{Limitations and Future Work}

\subsection{Trade-off between Candidate Diversity and Interaction Efficiency}
\label{sec:overall-discussion:trade-off}
Some participants initially tried to adjust the exact parameter values but were not satisfied with the outcomes. 
For example: \textit{``When I asked for something like 'set Objective1 to 80 or higher,' it didn't seem to match exactly.'' (Study 1, P8)}
Highly specific requests tended to fail when no available candidate satisfied them. 

Over time, they shifted to giving broader requests and letting the AI decide, which often led to better results and a stronger sense that the system understood their intent:
\textit{``Initially, I said 'raise parameter XX,' but later I just described the result I wanted and let the AI choose. That worked better.''} (Study 1, P18).
Such broad requests (\eg, ``improve overall performance'') often align with at least one BO candidate, which helped participants feel that the system understood their intent. 

Since we used a batch size of $q=8$ in both studies, increasing $q$ might provide more diverse candidates and better align with specific user requests. 
However, a larger batch size may require more time to sample candidates, which could negatively affect the efficiency of the design exploration.
This highlights a trade-off between diversity and interaction efficiency. 
Determining the optimal batch size that balances this trade-off remains an open question and is left for future work.

\subsection{Improving Early-Stage Accuracy of AI Suggestions}
\label{sec:discussion:early_stage_accuracy}

Because the GP has limited predictive accuracy in the early stages, batch BO may suggest candidates that do not match the designer's requests. 
In the initial phase of the design task, some participants felt that the system proposed off-target solutions: 
\textit{``Calling them 'off-target' might sound harsh, but initially I saw values drifting to the lower-left region. Still, since it won't improve without trying, I didn't think it was ignoring my requests. Overall, it listened to about half of them.'' (Study 2, P3)}
As more data accumulates, the GP becomes more accurate, and later-stage proposals better reflect the designer's request. 
Many participants noted this improvement over time, highlighting a difference in responsiveness depending on the phase of exploration. 
For example, one participant stated, \textit{``By the later stages, after the AI had accumulated more training data, its accuracy improved.'' (Study 1, P14)}

An important direction for future work is improving early-stage performance.
For example, using more initial seed points could boost early performance and give designers a stronger sense that ``the system is listening.''

\subsection{Leveraging Predictive Confidence in Interaction}
One promising future direction is to enhance interaction by better utilizing the system's confidence about the performance prediction and design proposals.
Currently, our system incorporates the BO's confidence (\ie, the predictive variance) into the LLM prompt.
As a result, the LLM occasionally generated relevant explanations such as ``the variance is small, so this prediction is reliable,'' even though it is not explicitly instructed to reason about uncertainty.
However, this approach does not guarantee consistent communication of uncertainty, which may lead to overconfidence or misunderstanding on the part of designers.
This issue is closely related to the broader challenge of ensuring appropriate reliance on AI---an important topic in the HCI community.

To address this, future work should explore interaction techniques that more effectively communicate uncertainty to designers.
This points to a clear direction: adapting the system's explanations based on its level of uncertainty.
For example, the LLM could be guided to generate more nuanced explanations, such as: ``This design is predicted to be effective, but the system has not explored similar designs in this area very much, so there may be uncertainty.''

\subsection{Applicability to Real-World Design Problems}

In our experiment, we used artificial and relatively simple interfaces to enable controlled evaluation of interaction.
An important question is whether our method can be applied to real-world design problems, where the parameter space tends to be higher-dimensional and performance depends on complex interactions among parameters.
We believe that, in such settings, the benefits of our approach become even greater.

As the design space becomes higher-dimensional, designers face significant challenges when manually specifying search ranges on a GUI.
Similarly, in complex problems where the relationship between parameters and multiple objectives is not intuitive, it is difficult to determine which parameter changes will lead to desirable outcomes.
In contrast, our method allows designers to provide collective instructions for high-dimensional problems, such as ``Adjust the layout-related parameters,'' and goal-based instructions for complex problems, such as ``Increase daily revenue.''
The flexibility of intervention that natural language provides plays a key role in keeping the designer's cognitive load low.
As the design space becomes more high-dimensional and complex, reducing cognitive load becomes critical and further enhances the conceptual advantage of our method.

Applying our method to high-dimensional design spaces poses challenges both inherent to Bayesian optimization---such as difficulties in surrogate modeling---and common to high-dimensional optimization in general, where the number of required iterations tends to grow rapidly.
While addressing these issues is beyond the scope of this work, they could be mitigated by recent advances in high-dimensional BO methods \cite{Hvarfner:ICML:2024} or by applying techniques such as dimensionality reduction.

That said, in real-world design practice involving iterative user testing, designers rarely adjust a large number of parameters at once.
This is because it becomes extremely difficult to identify the causal relationship between parameter changes and the resulting outcomes.
Instead, designers typically select and focus on a specific subset of parameters---usually 2 to 5 at a time \cite{Chan:CHI:22,Liao:CHI:24,Liao:PC:23}---and iterate based on the feedback.
Our current system aligns well with this practice.
Furthermore, future work could explore ways to semi-automatically select subsets of interest based on natural language requests, such as ``I want to focus on optimizing the layout parameters.''
Therefore, even as the total number of parameters increases, our approach can remain valuable when used in this practical manner of targeting specific subsets.

\subsection{Generalization to Unknown Design Spaces}

While our method appears promising applicability to real-world design tasks, its effectiveness in unfamiliar design spaces where the LLM has no prior domain knowledge remains an open question.
Nonetheless, we believe our method holds strong potential for generalization to such design spaces, owing to a hybrid approach that combines two key mechanisms.

First, our method leverages statistical indicators from the BO process. The prompt includes each candidate's predictive mean and variance, which serve as indicators of performance and uncertainty, respectively. This information enables the LLM to make data-driven decisions and balance the trade-off between exploitation and exploration without relying on domain-specific knowledge.

Second, the LLM uses its in-context learning ability to progressively learn the relationship between parameters and their outcomes from the evaluation history in the prompt. As more data is collected, the LLM can make increasingly informed selections, even if it starts with no understanding of this relationship. The combination of BO's data-driven guidance and the LLM's adaptive learning allows our method to generalize across new design spaces without requiring domain-specific knowledge.

Future work includes exploring the application of our method to design tasks unfamiliar to the LLM, such as tuning haptic feedback patterns on novel devices, to validate its generalization capability.

% !TEX root = ../paper.tex

\section{Conclusion}

Our goal is to facilitate cooperative design optimization through natural language interaction.
For this purpose, we propose a novel method that integrates BO with LLMs, 
where the use of LLMs allows designers to provide flexible, natural language requests to the optimization system and also receive natural language reasoning of the system's proposal.
Specifically, this is enabled by sampling multiple candidate points using a batch BO technique and then selecting the most suitable one from the candidates by interpreting the designer's request through an LLM.
Our user studies show that natural language-based cooperative optimization can provide higher user agency than a system-led method and shows promising optimization performance compared to manual design.
Moreover, it offers similar performance to an existing cooperative approach, with the added benefit of reduced cognitive load.
These results suggest that natural language interaction is an effective means to balance algorithmic efficiency with human control.
Overall, our findings highlight the potential of a new design framework that supports more seamless human--AI collaboration.

\begin{acks}
This work was supported by JST Moonshot R\&D Program Grant Number JPMJMS2236.
\end{acks}

\bibliographystyle{ACM-Reference-Format}
\bibliography{sample-base}


\begin{thebibliography}{36}

%%% the Web address.

\ifx \showCODEN    \undefined \def \showCODEN     #1{\unskip}     \fi
\ifx \showDOI      \undefined \def \showDOI       #1{#1}\fi
\ifx \showISBNx    \undefined \def \showISBNx     #1{\unskip}     \fi
\ifx \showISBNxiii \undefined \def \showISBNxiii  #1{\unskip}     \fi
\ifx \showISSN     \undefined \def \showISSN      #1{\unskip}     \fi
\ifx \showLCCN     \undefined \def \showLCCN      #1{\unskip}     \fi
\ifx \shownote     \undefined \def \shownote      #1{#1}          \fi
\ifx \showarticletitle \undefined \def \showarticletitle #1{#1}   \fi
\ifx \showURL      \undefined \def \showURL       {\relax}        \fi
% invisible to TeX
\providecommand\bibfield[2]{#2}
\providecommand\bibinfo[2]{#2}
\providecommand\natexlab[1]{#1}
\providecommand\showeprint[2][]{arXiv:#2}

\bibitem[AhmadiTeshnizi et~al\mbox{.}(2024)]%
        {AhmadiTeshnizi:ICML:2024}
\bibfield{author}{\bibinfo{person}{Ali AhmadiTeshnizi}, \bibinfo{person}{Wenzhi Gao}, {and} \bibinfo{person}{Madeleine Udell}.} \bibinfo{year}{2024}\natexlab{}.
\newblock \showarticletitle{OptiMUS: scalable optimization modeling with (MI)LP solvers and large language models}. In \bibinfo{booktitle}{\emph{Proceedings of the 41st International Conference on Machine Learning}} (Vienna, Austria) \emph{(\bibinfo{series}{ICML'24})}. \bibinfo{publisher}{JMLR.org}, Article \bibinfo{articleno}{24}, \bibinfo{numpages}{20}~pages.
\newblock


\bibitem[Balandat et~al\mbox{.}(2020)]%
        {Balandat:2020:botorch}
\bibfield{author}{\bibinfo{person}{Maximilian Balandat}, \bibinfo{person}{Brian Karrer}, \bibinfo{person}{Daniel~R. Jiang}, \bibinfo{person}{Samuel Daulton}, \bibinfo{person}{Benjamin Letham}, \bibinfo{person}{Andrew~Gordon Wilson}, {and} \bibinfo{person}{Eytan Bakshy}.} \bibinfo{year}{2020}\natexlab{}.
\newblock \showarticletitle{{BoTorch: A Framework for Efficient Monte-Carlo Bayesian Optimization}}. In \bibinfo{booktitle}{\emph{Advances in Neural Information Processing Systems 33}}.
\newblock
\urldef\tempurl%
\url{http://arxiv.org/abs/1910.06403}
\showURL{%
\tempurl}


\bibitem[Bi et~al\mbox{.}(2014)]%
        {Bi:CHI:14}
\bibfield{author}{\bibinfo{person}{Xiaojun Bi}, \bibinfo{person}{Tom Ouyang}, {and} \bibinfo{person}{Shumin Zhai}.} \bibinfo{year}{2014}\natexlab{}.
\newblock \showarticletitle{Both complete and correct? multi-objective optimization of touchscreen keyboard}. In \bibinfo{booktitle}{\emph{Proceedings of the SIGCHI Conference on Human Factors in Computing Systems}} (Toronto, Ontario, Canada) \emph{(\bibinfo{series}{CHI '14})}. \bibinfo{publisher}{Association for Computing Machinery}, \bibinfo{address}{New York, NY, USA}, \bibinfo{pages}{2297--2306}.
\newblock
\showISBNx{9781450324731}
\urldef\tempurl%
\url{https://doi.org/10.1145/2556288.2557414}
\showDOI{\tempurl}


\bibitem[Chan et~al\mbox{.}(2022)]%
        {Chan:CHI:22}
\bibfield{author}{\bibinfo{person}{Liwei Chan}, \bibinfo{person}{Yi-Chi Liao}, \bibinfo{person}{George~B Mo}, \bibinfo{person}{John~J Dudley}, \bibinfo{person}{Chun-Lien Cheng}, \bibinfo{person}{Per~Ola Kristensson}, {and} \bibinfo{person}{Antti Oulasvirta}.} \bibinfo{year}{2022}\natexlab{}.
\newblock \showarticletitle{Investigating Positive and Negative Qualities of Human-in-the-Loop Optimization for Designing Interaction Techniques}. In \bibinfo{booktitle}{\emph{Proceedings of the 2022 CHI Conference on Human Factors in Computing Systems}} (New Orleans, LA, USA) \emph{(\bibinfo{series}{CHI '22})}. \bibinfo{publisher}{Association for Computing Machinery}, \bibinfo{address}{New York, NY, USA}, Article \bibinfo{articleno}{112}, \bibinfo{numpages}{14}~pages.
\newblock
\showISBNx{9781450391573}
\urldef\tempurl%
\url{https://doi.org/10.1145/3491102.3501850}
\showDOI{\tempurl}


\bibitem[Ciss{\'e} et~al\mbox{.}(2025)]%
        {cisse:arXiv:2025}
\bibfield{author}{\bibinfo{person}{Abdoulatif Ciss{\'e}}, \bibinfo{person}{Xenophon Evangelopoulos}, \bibinfo{person}{Vladimir~V Gusev}, {and} \bibinfo{person}{Andrew~I Cooper}.} \bibinfo{year}{2025}\natexlab{}.
\newblock \showarticletitle{Language-Based Bayesian Optimization Research Assistant (BORA)}.
\newblock \bibinfo{journal}{\emph{arXiv preprint arXiv:2501.16224}} (\bibinfo{year}{2025}).
\newblock


\bibitem[Dafoe et~al\mbox{.}(2021)]%
        {Dafoe:Nature:21}
\bibfield{author}{\bibinfo{person}{Allan Dafoe}, \bibinfo{person}{Yoram Bachrach}, \bibinfo{person}{Gillian Hadfield}, \bibinfo{person}{Eric Horvitz}, \bibinfo{person}{Kate Larson}, {and} \bibinfo{person}{Thore Graepel}.} \bibinfo{year}{2021}\natexlab{}.
\newblock \showarticletitle{Cooperative AI: Machines Must Learn to Find Common Ground}.
\newblock \bibinfo{journal}{\emph{Nature}}  \bibinfo{volume}{593} (\bibinfo{date}{May} \bibinfo{year}{2021}), \bibinfo{pages}{33--36}.
\newblock
\urldef\tempurl%
\url{https://doi.org/10.1038/d41586-021-01170-0}
\showDOI{\tempurl}


\bibitem[Daulton et~al\mbox{.}(2023)]%
        {Daulton:NIPS:23}
\bibfield{author}{\bibinfo{person}{Samuel Daulton}, \bibinfo{person}{Sebastian Ament}, \bibinfo{person}{David Eriksson}, \bibinfo{person}{Maximilian Balandat}, {and} \bibinfo{person}{Eytan Bakshy}.} \bibinfo{year}{2023}\natexlab{}.
\newblock \showarticletitle{Unexpected improvements to expected improvement for Bayesian optimization}. In \bibinfo{booktitle}{\emph{Proceedings of the 37th International Conference on Neural Information Processing Systems}} (New Orleans, LA, USA) \emph{(\bibinfo{series}{NIPS '23})}. \bibinfo{publisher}{Curran Associates Inc.}, \bibinfo{address}{Red Hook, NY, USA}, Article \bibinfo{articleno}{904}, \bibinfo{numpages}{36}~pages.
\newblock


\bibitem[Dudley et~al\mbox{.}(2019)]%
        {Dudley:CHI:19}
\bibfield{author}{\bibinfo{person}{John~J. Dudley}, \bibinfo{person}{Jason~T. Jacques}, {and} \bibinfo{person}{Per~Ola Kristensson}.} \bibinfo{year}{2019}\natexlab{}.
\newblock \showarticletitle{Crowdsourcing Interface Feature Design with Bayesian Optimization}. In \bibinfo{booktitle}{\emph{Proceedings of the 2019 CHI Conference on Human Factors in Computing Systems}} (Glasgow, Scotland Uk) \emph{(\bibinfo{series}{CHI '19})}. \bibinfo{publisher}{Association for Computing Machinery}, \bibinfo{address}{New York, NY, USA}, \bibinfo{pages}{1--12}.
\newblock
\showISBNx{9781450359702}
\urldef\tempurl%
\url{https://doi.org/10.1145/3290605.3300482}
\showDOI{\tempurl}


\bibitem[Frazier(2018)]%
        {Frazier:arxiv:2018}
\bibfield{author}{\bibinfo{person}{Peter~I Frazier}.} \bibinfo{year}{2018}\natexlab{}.
\newblock \showarticletitle{A tutorial on Bayesian optimization}.
\newblock \bibinfo{journal}{\emph{arXiv preprint arXiv:1807.02811}} (\bibinfo{year}{2018}).
\newblock


\bibitem[Hao et~al\mbox{.}(2024)]%
        {Hao:SEC:2024}
\bibfield{author}{\bibinfo{person}{Hao Hao}, \bibinfo{person}{Xiaoqun Zhang}, {and} \bibinfo{person}{Aimin Zhou}.} \bibinfo{year}{2024}\natexlab{}.
\newblock \showarticletitle{Large language models as surrogate models in evolutionary algorithms: A preliminary study}.
\newblock \bibinfo{journal}{\emph{Swarm and Evolutionary Computation}}  \bibinfo{volume}{91} (\bibinfo{year}{2024}), \bibinfo{pages}{101741}.
\newblock
\showISSN{2210-6502}
\urldef\tempurl%
\url{https://doi.org/10.1016/j.swevo.2024.101741}
\showDOI{\tempurl}


\bibitem[Hart and Staveland(1988)]%
        {Hart:NASATLX:1988}
\bibfield{author}{\bibinfo{person}{Sandra~G. Hart} {and} \bibinfo{person}{Lowell~E. Staveland}.} \bibinfo{year}{1988}\natexlab{}.
\newblock \showarticletitle{Development of NASA-TLX (Task Load Index): Results of Empirical and Theoretical Research}.
\newblock In \bibinfo{booktitle}{\emph{Human Mental Workload}}, \bibfield{editor}{\bibinfo{person}{Peter~A. Hancock} {and} \bibinfo{person}{Najmedin Meshkati}} (Eds.). \bibinfo{series}{Advances in Psychology}, Vol.~\bibinfo{volume}{52}. \bibinfo{publisher}{North-Holland}, \bibinfo{pages}{139--183}.
\newblock
\showISSN{0166-4115}
\urldef\tempurl%
\url{https://doi.org/10.1016/S0166-4115(08)62386-9}
\showDOI{\tempurl}


\bibitem[Huang et~al\mbox{.}(2025)]%
        {Chenyu:arxiv:2025}
\bibfield{author}{\bibinfo{person}{Chenyu Huang}, \bibinfo{person}{Zhengyang Tang}, \bibinfo{person}{Shixi Hu}, \bibinfo{person}{Ruoqing Jiang}, \bibinfo{person}{Xin Zheng}, \bibinfo{person}{Dongdong Ge}, \bibinfo{person}{Benyou Wang}, {and} \bibinfo{person}{Zizhuo Wang}.} \bibinfo{year}{2025}\natexlab{}.
\newblock \bibinfo{title}{ORLM: A Customizable Framework in Training Large Models for Automated Optimization Modeling}.
\newblock
\newblock
\showeprint[arxiv]{2405.17743}~[cs.CL]
\urldef\tempurl%
\url{https://arxiv.org/abs/2405.17743}
\showURL{%
\tempurl}


\bibitem[Huang et~al\mbox{.}(2024)]%
        {Huang:SEC:2024}
\bibfield{author}{\bibinfo{person}{Sen Huang}, \bibinfo{person}{Kaixiang Yang}, \bibinfo{person}{Sheng Qi}, {and} \bibinfo{person}{Rui Wang}.} \bibinfo{year}{2024}\natexlab{}.
\newblock \showarticletitle{When large language model meets optimization}.
\newblock \bibinfo{journal}{\emph{Swarm and Evolutionary Computation}}  \bibinfo{volume}{90} (\bibinfo{year}{2024}), \bibinfo{pages}{101663}.
\newblock
\showISSN{2210-6502}
\urldef\tempurl%
\url{https://doi.org/10.1016/j.swevo.2024.101663}
\showDOI{\tempurl}


\bibitem[Hvarfner et~al\mbox{.}(2024)]%
        {Hvarfner:ICML:2024}
\bibfield{author}{\bibinfo{person}{Carl Hvarfner}, \bibinfo{person}{Erik~Orm Hellsten}, {and} \bibinfo{person}{Luigi Nardi}.} \bibinfo{year}{2024}\natexlab{}.
\newblock \showarticletitle{Vanilla {B}ayesian Optimization Performs Great in High Dimensions}. In \bibinfo{booktitle}{\emph{Proceedings of the 41st International Conference on Machine Learning}} \emph{(\bibinfo{series}{Proceedings of Machine Learning Research}, Vol.~\bibinfo{volume}{235})}, \bibfield{editor}{\bibinfo{person}{Ruslan Salakhutdinov}, \bibinfo{person}{Zico Kolter}, \bibinfo{person}{Katherine Heller}, \bibinfo{person}{Adrian Weller}, \bibinfo{person}{Nuria Oliver}, \bibinfo{person}{Jonathan Scarlett}, {and} \bibinfo{person}{Felix Berkenkamp}} (Eds.). \bibinfo{publisher}{PMLR}, \bibinfo{pages}{20793--20817}.
\newblock
\urldef\tempurl%
\url{https://proceedings.mlr.press/v235/hvarfner24a.html}
\showURL{%
\tempurl}


\bibitem[Khajah et~al\mbox{.}(2016)]%
        {Khajah:CHI:16}
\bibfield{author}{\bibinfo{person}{Mohammad~M. Khajah}, \bibinfo{person}{Brett~D. Roads}, \bibinfo{person}{Robert~V. Lindsey}, \bibinfo{person}{Yun-En Liu}, {and} \bibinfo{person}{Michael~C. Mozer}.} \bibinfo{year}{2016}\natexlab{}.
\newblock \showarticletitle{Designing Engaging Games Using Bayesian Optimization}. In \bibinfo{booktitle}{\emph{Proceedings of the 2016 CHI Conference on Human Factors in Computing Systems}} (San Jose, California, USA) \emph{(\bibinfo{series}{CHI '16})}. \bibinfo{publisher}{Association for Computing Machinery}, \bibinfo{address}{New York, NY, USA}, \bibinfo{pages}{5571--5582}.
\newblock
\showISBNx{9781450333627}
\urldef\tempurl%
\url{https://doi.org/10.1145/2858036.2858253}
\showDOI{\tempurl}


\bibitem[Koyama and Goto(2022)]%
        {Koyama:UIST:22}
\bibfield{author}{\bibinfo{person}{Yuki Koyama} {and} \bibinfo{person}{Masataka Goto}.} \bibinfo{year}{2022}\natexlab{}.
\newblock \showarticletitle{{BO as Assistant}: Using {Bayesian} Optimization for Asynchronously Generating Design Suggestions}. In \bibinfo{booktitle}{\emph{Proceedings of the 35th Annual ACM Symposium on User Interface Software and Technology}} (Bend, OR, USA) \emph{(\bibinfo{series}{UIST '22})}. \bibinfo{publisher}{Association for Computing Machinery}, \bibinfo{address}{New York, NY, USA}, Article \bibinfo{articleno}{77}, \bibinfo{numpages}{14}~pages.
\newblock
\showISBNx{9781450393201}
\urldef\tempurl%
\url{https://doi.org/10.1145/3526113.3545664}
\showDOI{\tempurl}


\bibitem[Koyama et~al\mbox{.}(2020)]%
        {Koyama:SIGGRAPH:20}
\bibfield{author}{\bibinfo{person}{Yuki Koyama}, \bibinfo{person}{Issei Sato}, {and} \bibinfo{person}{Masataka Goto}.} \bibinfo{year}{2020}\natexlab{}.
\newblock \showarticletitle{Sequential gallery for interactive visual design optimization}.
\newblock \bibinfo{journal}{\emph{ACM Trans. Graph.}} \bibinfo{volume}{39}, \bibinfo{number}{4}, Article \bibinfo{articleno}{88} (\bibinfo{date}{aug} \bibinfo{year}{2020}), \bibinfo{numpages}{12}~pages.
\newblock
\showISSN{0730-0301}
\urldef\tempurl%
\url{https://doi.org/10.1145/3386569.3392444}
\showDOI{\tempurl}


\bibitem[Koyama et~al\mbox{.}(2017)]%
        {Koyama:SIGGRAPH:17}
\bibfield{author}{\bibinfo{person}{Yuki Koyama}, \bibinfo{person}{Issei Sato}, \bibinfo{person}{Daisuke Sakamoto}, {and} \bibinfo{person}{Takeo Igarashi}.} \bibinfo{year}{2017}\natexlab{}.
\newblock \showarticletitle{Sequential line search for efficient visual design optimization by crowds}.
\newblock \bibinfo{journal}{\emph{ACM Trans. Graph.}} \bibinfo{volume}{36}, \bibinfo{number}{4}, Article \bibinfo{articleno}{48} (\bibinfo{date}{jul} \bibinfo{year}{2017}), \bibinfo{numpages}{11}~pages.
\newblock
\showISSN{0730-0301}
\urldef\tempurl%
\url{https://doi.org/10.1145/3072959.3073598}
\showDOI{\tempurl}


\bibitem[Lawless et~al\mbox{.}(2024)]%
        {Lawless:TIIS:2024}
\bibfield{author}{\bibinfo{person}{Connor Lawless}, \bibinfo{person}{Jakob Schoeffer}, \bibinfo{person}{Lindy Le}, \bibinfo{person}{Kael Rowan}, \bibinfo{person}{Shilad Sen}, \bibinfo{person}{Cristina St.~Hill}, \bibinfo{person}{Jina Suh}, {and} \bibinfo{person}{Bahareh Sarrafzadeh}.} \bibinfo{year}{2024}\natexlab{}.
\newblock \showarticletitle{“I Want It That Way”: Enabling Interactive Decision Support Using Large Language Models and Constraint Programming}.
\newblock \bibinfo{journal}{\emph{ACM Trans. Interact. Intell. Syst.}} \bibinfo{volume}{14}, \bibinfo{number}{3}, Article \bibinfo{articleno}{22} (\bibinfo{date}{Sept.} \bibinfo{year}{2024}), \bibinfo{numpages}{33}~pages.
\newblock
\showISSN{2160-6455}
\urldef\tempurl%
\url{https://doi.org/10.1145/3685053}
\showDOI{\tempurl}


\bibitem[Li et~al\mbox{.}(2023)]%
        {Li:arxiv:2023}
\bibfield{author}{\bibinfo{person}{Beibin Li}, \bibinfo{person}{Konstantina Mellou}, \bibinfo{person}{Bo Zhang}, \bibinfo{person}{Jeevan Pathuri}, {and} \bibinfo{person}{Ishai Menache}.} \bibinfo{year}{2023}\natexlab{}.
\newblock \bibinfo{title}{Large Language Models for Supply Chain Optimization}.
\newblock
\newblock
\showeprint[arxiv]{2307.03875}~[cs.AI]
\urldef\tempurl%
\url{https://arxiv.org/abs/2307.03875}
\showURL{%
\tempurl}


\bibitem[Liao et~al\mbox{.}(2024)]%
        {Liao:CHI:24}
\bibfield{author}{\bibinfo{person}{Yi-Chi Liao}, \bibinfo{person}{Ruta Desai}, \bibinfo{person}{Alec~M Pierce}, \bibinfo{person}{Krista~E. Taylor}, \bibinfo{person}{Hrvoje Benko}, \bibinfo{person}{Tanya~R. Jonker}, {and} \bibinfo{person}{Aakar Gupta}.} \bibinfo{year}{2024}\natexlab{}.
\newblock \showarticletitle{A Meta-Bayesian Approach for Rapid Online Parametric Optimization for Wrist-based Interactions}. In \bibinfo{booktitle}{\emph{Proceedings of the 2024 CHI Conference on Human Factors in Computing Systems}} (Honolulu, HI, USA) \emph{(\bibinfo{series}{CHI '24})}. \bibinfo{publisher}{Association for Computing Machinery}, \bibinfo{address}{New York, NY, USA}, Article \bibinfo{articleno}{410}, \bibinfo{numpages}{38}~pages.
\newblock
\showISBNx{9798400703300}
\urldef\tempurl%
\url{https://doi.org/10.1145/3613904.3642071}
\showDOI{\tempurl}


\bibitem[Liao et~al\mbox{.}(2023)]%
        {Liao:PC:23}
\bibfield{author}{\bibinfo{person}{Yi-Chi Liao}, \bibinfo{person}{John~J. Dudley}, \bibinfo{person}{George~B. Mo}, \bibinfo{person}{Chun-Lien Cheng}, \bibinfo{person}{Liwei Chan}, \bibinfo{person}{Antti Oulasvirta}, {and} \bibinfo{person}{Per~Ola Kristensson}.} \bibinfo{year}{2023}\natexlab{}.
\newblock \showarticletitle{Interaction Design With Multi-Objective Bayesian Optimization}.
\newblock \bibinfo{journal}{\emph{IEEE Pervasive Computing}} \bibinfo{volume}{22}, \bibinfo{number}{1} (\bibinfo{year}{2023}), \bibinfo{pages}{29--38}.
\newblock
\urldef\tempurl%
\url{https://doi.org/10.1109/MPRV.2022.3230597}
\showDOI{\tempurl}


\bibitem[Liu et~al\mbox{.}(2023)]%
        {Liu:TIIS:2023}
\bibfield{author}{\bibinfo{person}{Jie Liu}, \bibinfo{person}{Kim Marriott}, \bibinfo{person}{Tim Dwyer}, {and} \bibinfo{person}{Guido Tack}.} \bibinfo{year}{2023}\natexlab{}.
\newblock \showarticletitle{Increasing User Trust in Optimisation through Feedback and Interaction}.
\newblock \bibinfo{journal}{\emph{ACM Trans. Comput.-Hum. Interact.}} \bibinfo{volume}{29}, \bibinfo{number}{5}, Article \bibinfo{articleno}{42} (\bibinfo{date}{Jan.} \bibinfo{year}{2023}), \bibinfo{numpages}{34}~pages.
\newblock
\showISSN{1073-0516}
\urldef\tempurl%
\url{https://doi.org/10.1145/3503461}
\showDOI{\tempurl}


\bibitem[Liu et~al\mbox{.}(2024)]%
        {Liu:ICLR:24}
\bibfield{author}{\bibinfo{person}{Tennison Liu}, \bibinfo{person}{Nicol{\'a}s Astorga}, \bibinfo{person}{Nabeel Seedat}, {and} \bibinfo{person}{Mihaela van~der Schaar}.} \bibinfo{year}{2024}\natexlab{}.
\newblock \showarticletitle{Large Language Models to Enhance Bayesian Optimization}. In \bibinfo{booktitle}{\emph{The Twelfth International Conference on Learning Representations}}.
\newblock
\urldef\tempurl%
\url{https://openreview.net/forum?id=OOxotBmGol}
\showURL{%
\tempurl}


\bibitem[Mahammadli and Ertekin(2024)]%
        {Kanan:arXiv:2024}
\bibfield{author}{\bibinfo{person}{Kanan Mahammadli} {and} \bibinfo{person}{Seyda Ertekin}.} \bibinfo{year}{2024}\natexlab{}.
\newblock \showarticletitle{Sequential large language model-based hyper-parameter optimization}.
\newblock \bibinfo{journal}{\emph{arXiv preprint arXiv:2410.20302}} (\bibinfo{year}{2024}).
\newblock


\bibitem[Meinhardt et~al\mbox{.}(2025)]%
        {Meinhardt:CHI:2025}
\bibfield{author}{\bibinfo{person}{Luca-Maxim Meinhardt}, \bibinfo{person}{Clara Schramm}, \bibinfo{person}{Pascal Jansen}, \bibinfo{person}{Mark Colley}, {and} \bibinfo{person}{Enrico Rukzio}.} \bibinfo{year}{2025}\natexlab{}.
\newblock \showarticletitle{Fly Away: Evaluating the Impact of Motion Fidelity on Optimized User Interface Design via Bayesian Optimization in Automated Urban Air Mobility Simulations}.
\newblock \bibinfo{journal}{\emph{arXiv preprint arXiv:2501.11829}} (\bibinfo{year}{2025}).
\newblock


\bibitem[Michailidis et~al\mbox{.}(2024)]%
        {Michailidis:CP:2024}
\bibfield{author}{\bibinfo{person}{Kostis Michailidis}, \bibinfo{person}{Dimos Tsouros}, {and} \bibinfo{person}{Tias Guns}.} \bibinfo{year}{2024}\natexlab{}.
\newblock \showarticletitle{{Constraint Modelling with LLMs Using In-Context Learning}}. In \bibinfo{booktitle}{\emph{30th International Conference on Principles and Practice of Constraint Programming (CP 2024)}} \emph{(\bibinfo{series}{Leibniz International Proceedings in Informatics (LIPIcs)}, Vol.~\bibinfo{volume}{307})}, \bibfield{editor}{\bibinfo{person}{Paul Shaw}} (Ed.). \bibinfo{publisher}{Schloss Dagstuhl -- Leibniz-Zentrum f{\"u}r Informatik}, \bibinfo{address}{Dagstuhl, Germany}, \bibinfo{pages}{20:1--20:27}.
\newblock
\showISBNx{978-3-95977-336-2}
\showISSN{1868-8969}
\urldef\tempurl%
\url{https://doi.org/10.4230/LIPIcs.CP.2024.20}
\showDOI{\tempurl}


\bibitem[Mo et~al\mbox{.}(2024)]%
        {Mo:TIIS:24}
\bibfield{author}{\bibinfo{person}{George Mo}, \bibinfo{person}{John Dudley}, \bibinfo{person}{Liwei Chan}, \bibinfo{person}{Yi-Chi Liao}, \bibinfo{person}{Antti Oulasvirta}, {and} \bibinfo{person}{Per~Ola Kristensson}.} \bibinfo{year}{2024}\natexlab{}.
\newblock \showarticletitle{Cooperative Multi-Objective Bayesian Design Optimization}.
\newblock \bibinfo{journal}{\emph{ACM Trans. Interact. Intell. Syst.}} \bibinfo{volume}{14}, \bibinfo{number}{2}, Article \bibinfo{articleno}{13} (\bibinfo{date}{jun} \bibinfo{year}{2024}), \bibinfo{numpages}{28}~pages.
\newblock
\showISSN{2160-6455}
\urldef\tempurl%
\url{https://doi.org/10.1145/3657643}
\showDOI{\tempurl}


\bibitem[Ramos et~al\mbox{.}(2023)]%
        {Ramos:arXiv:23}
\bibfield{author}{\bibinfo{person}{Mayk~Caldas Ramos}, \bibinfo{person}{Shane~S Michtavy}, \bibinfo{person}{Marc~D Porosoff}, {and} \bibinfo{person}{Andrew~D White}.} \bibinfo{year}{2023}\natexlab{}.
\newblock \showarticletitle{Bayesian optimization of catalysts with in-context learning}.
\newblock \bibinfo{journal}{\emph{arXiv [physics.chem-ph]}} (\bibinfo{date}{April} \bibinfo{year}{2023}).
\newblock


\bibitem[Rankovi{\'c} and Schwaller(2023)]%
        {rankovic:NeurIPS:2023}
\bibfield{author}{\bibinfo{person}{Bojana Rankovi{\'c}} {and} \bibinfo{person}{Philippe Schwaller}.} \bibinfo{year}{2023}\natexlab{}.
\newblock \showarticletitle{BoChemian: Large Language Model Embeddings for Bayesian Optimization of Chemical Reactions}. In \bibinfo{booktitle}{\emph{NeurIPS 2023 Workshop on Adaptive Experimental Design and Active Learning in the Real World}}.
\newblock
\urldef\tempurl%
\url{https://openreview.net/forum?id=A1RVn1m3J3}
\showURL{%
\tempurl}


\bibitem[Roesler* et~al\mbox{.}(2022)]%
        {Roesler:HFESAM2022}
\bibfield{author}{\bibinfo{person}{Eileen Roesler*}, \bibinfo{person}{Tobias Rieger*}, {and} \bibinfo{person}{Dietrich Manzey}.} \bibinfo{year}{2022}\natexlab{}.
\newblock \showarticletitle{Trust towards Human vs. Automated Agents: Using a Multidimensional Trust Questionnaire to Assess The Role of Performance, Utility, Purpose, and Transparency}.
\newblock \bibinfo{journal}{\emph{Proceedings of the Human Factors and Ergonomics Society Annual Meeting}} \bibinfo{volume}{66}, \bibinfo{number}{1} (\bibinfo{year}{2022}), \bibinfo{pages}{2047--2051}.
\newblock
\urldef\tempurl%
\url{https://doi.org/10.1177/1071181322661065}
\showDOI{\tempurl}


\bibitem[Shahriari et~al\mbox{.}(2016)]%
        {Shahriari:ProcIEEE:16}
\bibfield{author}{\bibinfo{person}{Bobak Shahriari}, \bibinfo{person}{Kevin Swersky}, \bibinfo{person}{Ziyu Wang}, \bibinfo{person}{Ryan~P. Adams}, {and} \bibinfo{person}{Nando de Freitas}.} \bibinfo{year}{2016}\natexlab{}.
\newblock \showarticletitle{Taking the Human Out of the Loop: A Review of Bayesian Optimization}.
\newblock \bibinfo{journal}{\emph{Proc. IEEE}} \bibinfo{volume}{104}, \bibinfo{number}{1} (\bibinfo{year}{2016}), \bibinfo{pages}{148--175}.
\newblock
\urldef\tempurl%
\url{https://doi.org/10.1109/JPROC.2015.2494218}
\showDOI{\tempurl}


\bibitem[Wang et~al\mbox{.}(2024)]%
        {Wang:IUI:24}
\bibfield{author}{\bibinfo{person}{Bryan Wang}, \bibinfo{person}{Yuliang Li}, \bibinfo{person}{Zhaoyang Lv}, \bibinfo{person}{Haijun Xia}, \bibinfo{person}{Yan Xu}, {and} \bibinfo{person}{Raj Sodhi}.} \bibinfo{year}{2024}\natexlab{}.
\newblock \showarticletitle{LAVE: LLM-Powered Agent Assistance and Language Augmentation for Video Editing}. In \bibinfo{booktitle}{\emph{Proceedings of the 29th International Conference on Intelligent User Interfaces}} (Greenville, SC, USA) \emph{(\bibinfo{series}{IUI '24})}. \bibinfo{publisher}{Association for Computing Machinery}, \bibinfo{address}{New York, NY, USA}, \bibinfo{pages}{699–714}.
\newblock
\showISBNx{9798400705083}
\urldef\tempurl%
\url{https://doi.org/10.1145/3640543.3645143}
\showDOI{\tempurl}


\bibitem[Yamamoto et~al\mbox{.}(2022)]%
        {Yamamoto:UIST:22}
\bibfield{author}{\bibinfo{person}{Kenta Yamamoto}, \bibinfo{person}{Yuki Koyama}, {and} \bibinfo{person}{Yoichi Ochiai}.} \bibinfo{year}{2022}\natexlab{}.
\newblock \showarticletitle{Photographic Lighting Design with Photographer-in-the-Loop {Bayesian} Optimization}. In \bibinfo{booktitle}{\emph{Proceedings of the 35th Annual ACM Symposium on User Interface Software and Technology}} (Bend, OR, USA) \emph{(\bibinfo{series}{UIST '22})}. \bibinfo{publisher}{Association for Computing Machinery}, \bibinfo{address}{New York, NY, USA}, Article \bibinfo{articleno}{92}, \bibinfo{numpages}{11}~pages.
\newblock
\showISBNx{9781450393201}
\urldef\tempurl%
\url{https://doi.org/10.1145/3526113.3545690}
\showDOI{\tempurl}


\bibitem[Yang et~al\mbox{.}(2024)]%
        {yang2024large}
\bibfield{author}{\bibinfo{person}{Chengrun Yang}, \bibinfo{person}{Xuezhi Wang}, \bibinfo{person}{Yifeng Lu}, \bibinfo{person}{Hanxiao Liu}, \bibinfo{person}{Quoc~V Le}, \bibinfo{person}{Denny Zhou}, {and} \bibinfo{person}{Xinyun Chen}.} \bibinfo{year}{2024}\natexlab{}.
\newblock \showarticletitle{Large Language Models as Optimizers}. In \bibinfo{booktitle}{\emph{The Twelfth International Conference on Learning Representations}}.
\newblock
\urldef\tempurl%
\url{https://openreview.net/forum?id=Bb4VGOWELI}
\showURL{%
\tempurl}


\bibitem[Yoshida et~al\mbox{.}(2024)]%
        {Yoshida:Access:24}
\bibfield{author}{\bibinfo{person}{Shigeo Yoshida}, \bibinfo{person}{Yuki Koyama}, {and} \bibinfo{person}{Yoshitaka Ushiku}.} \bibinfo{year}{2024}\natexlab{}.
\newblock \showarticletitle{Toward AI-Mediated Avatar-Based Telecommunication: Investigating Visual Impression of Switching Between User- and AI-Controlled Avatars in Video Chat}.
\newblock \bibinfo{journal}{\emph{IEEE Access}}  \bibinfo{volume}{12} (\bibinfo{year}{2024}), \bibinfo{pages}{113372--113383}.
\newblock
\urldef\tempurl%
\url{https://doi.org/10.1109/ACCESS.2024.3441233}
\showDOI{\tempurl}


\end{thebibliography}

\appendix

% !TEX root = ../paper.tex

\section{Prompt}
\label{sec:appendix:prompt}

\figselectingprompts

\autoref{fig:overview_of_prompt} shows an overview of the prompt used to instruct the LLM during the optimization process.
Complete prompts for reproducibility are provided in the supplementary material.

\section{Technical Validation of Natural-Language Guidance}
\label{sec:appendix:tech-eval}

\figappendixtecheval

The proposed technique allows users to guide the optimization behavior through natural language.
We conducted a simple technical assessment to see how well this technique works.

We initialized the optimization with five randomly generated candidate points.
Over the next 15 iterations, the optimization process was guided using the proposed technique.
For this evaluation, we adopted a \emph{simulation} approach to emulate natural language requests by human designers; that is, we inserted artificial requests and then saw how the optimization behavior changes according to the requests.
Specifically, during the 15 iterations, we inserted the following two request prompts alternately: (1) ``Please propose parameters that increase Objective 1'' and (2) ``Please propose parameters that increase Objective 2.'' 
We then evaluated the returned parameters using the synthetic test functions (\autoref{sec:appendix:tasks}) used in Study 1 and Study 2.
This was repeated five times with different random seeds.

As shown in each subplot of \autoref{fig:tech_eval}, the results of the requests focusing on Objective 1 (red dots) tend to cluster at higher values of Objective 1 (right side of the plot), while those focusing on Objective 2 (blue dots) tend to cluster at higher values of Objective 1 (top side of the plot).
This provides visual evidence that the requests could successfully guide the optimization behavior.

Finally, to measure how distinctly these clusters separate in the parameter space, we calculated the Euclidean distances between their centroids.
The values were $0.215 \pm 0.004$ for App 1, $0.206 \pm 0.000$ for App 2, and $0.187 \pm 0.006$ for App 3.
Note that if the natural language-based guidance has no effect, these values should be closer to zero.
This indicates that alternating between Objective 1 and Objective 2 successfully steers the system toward different regions of the search space.

\section{Target Tasks and Synthetic Test Functions in User Study}
\label{sec:appendix:tasks}

\autoref{evaluation:parameters_and_objectives} shows the details of the design parameters and objectives in the user study tasks.

\autoref{evaluation:parameters_for_evaluation_function} lists the hyperparameters of the test functions for the three applications and the tutorial.
These hyperparameters were designed by Mo \etal\ to have a coherent behavior, and we adopted the same hyperparameters (\ie, increasing the density of notification frequency increases daily revenue but decreases user rating).

\begin{table*}[t]
    \centering
    \caption{Design parameters, objectives, and ranges for each of the applications (Apps 1--3) and tutorial (T).}
    \label{evaluation:parameters_and_objectives}
    \begin{tabular}{p{0.5cm} p{1.65cm} p{1.65cm} p{1.65cm} p{1.65cm} p{1.65cm} p{1.65cm} p{1.65cm}}
        \toprule
        \textbf{App} & \textbf{$x_1$} & \textbf{$x_2$} & \textbf{$x_3$} & \textbf{$x_4$} & \textbf{$x_5$} & \textbf{$f_1$} & \textbf{$f_2$} \\
            \midrule
            1 & Density of ads- [0,1] & Notification frequency [0,2] per hour & Personalization rate of content- [0,1] &  Moderation rate of content- [0,1] & Refresh time of content [0,20] minutes & Daily revenue- [0, 20] thousands USD & User rating- [0, 5] \\
            \midrule
            2 & Question categories- [5,50] & Refresh time of content [0,1000] & Length of question preview [0, 500] characters & Max number of question tags [1,10] & Threshold activity rating for user to answer questions [0,5] & Answering rate of questions- [0,2] per minute & Questions Answered- [0,100] \\
            \midrule
            3 & Location icon transparency - [0.5, 1] & Cursor distance for restaurant to show - [5, 50] & Location icon size - [1, 10] & Description box size - [10, 50] & Restaurant name text size - [10, 30] & Average speed to find restaurants- [0, 2] per minute & Accuracy in finding all restaurants- [0, 100] \\
            \midrule
            T & Force to register contact on screen - [10, 100]$N$ & Area to register contact on screen - [0.5, 3.0]$cm^2$ & - & - & - &Average target hit speed - [0, 3]per second & Accuracy of hitting targets- [0, 100]\% \\
        \bottomrule
    \end{tabular}
\end{table*}

\begin{table*}[ht]
    \centering
    \caption{Parameters $\mathbf{a}_1, \mathbf{a}_2, \mathbf{b}_1, \mathbf{b}_2, c_1, c_2$ for the synthetic test functions used in the applications (Apps 1--3) and tutorial (T).}
    \label{evaluation:parameters_for_evaluation_function}
    \begin{tabular}{p{0.5cm} p{2.9cm} p{2.7cm} p{2.7cm} p{2.7cm} p{0.5cm} p{0.5cm}}
        \toprule
        \textbf{App} & \textbf{$\mathbf{a}_1$} & \textbf{$\mathbf{a}_2$} & \textbf{$\mathbf{b}_1$} & \textbf{$\mathbf{b}_2$} & \textbf{$c_1$} & \textbf{$c_2$} \\
        \midrule
        1 & $[0.9,0.3,0.8,0.25,0.25]$ & $[0.3,0.35,1.1,0.75,0.3]$ & $[0.9,0.4,1.3,0.7,0.4]$ & $[1.0,0.6,1.2,0.5,0.4]$ & 0.7 & 0.8 \\
        \midrule
        2 & $[-0.1,0.25,0.7,0.7,0.65]$ & $[0.2,0.75,0.75,0.1,0.7]$ & $[1.2,0.5,0.4,1.0,0.6]$ & $[1.3,0.7,0.4,0.9,0.4]$ & 0.8 & 0.7 \\
        \midrule
        3 & $[1.1,0.75,0.35,0.3,0.3]$ & $[0.8,0.25,0.3,0.9,0.25]$ & $[1.2,0.5,0.6,1.0,0.4]$ & $[1.3,0.7,0.4,0.9,0.4]$ & 0.7 & 0.8 \\
        \midrule
        T & $[0.3,0.35]$ & $[0.7,0.65]$ & $[1.0,0.8]$ & $[1.2,0.9]$ & 0.7 & 0.8 \\
        \bottomrule
    \end{tabular}
\end{table*}

\section{Questionnaire and Detailed Results from User Study 1}
\label{sec:appendix_questionnaire_and_detailed_results_from_user_study_1}

\subsection{Questionnaire}
\subsubsection{Agency-Related Questionnaire}
\label{sec:appendix_user_study_1_agency_related_questionnaire}
\figappendixuserstudyfirstagencyboxplot
\figappendixuserstudyfirstdesignspaceunderstandingboxplot
\autoref{fig:user_study_first_agency_boxplot} presents the boxplots for Q1, Q2, and Q3, providing a visual comparison of the results across conditions. All items (Q1, Q2, Q3, and the Agency Score) showed significant differences among the three conditions (Designer-led, BO-led, and Cooperative (Natural Language)) based on Friedman tests (Q1: $p=0.001$; Q2: $p=0.001$; Q3: $p=0.050$; Agency Score: $p=0.001$). We then conducted post hoc Wilcoxon signed-rank tests with Bonferroni corrections, as summarized below.

For Q1, we found significant differences between the Designer-led and BO-led conditions ($p.\textrm{adj}=0.001$) and between the BO-led and Cooperative (Natural Language) conditions ($p.\textrm{adj}=0.001$). In contrast, there was no significant difference between the Designer-led and Cooperative (Natural Language) conditions ($p.\textrm{adj}=0.571$). Q2 followed a similar pattern, with significant differences between the Designer-led and BO-led conditions ($p.\textrm{adj}=0.003$) and between the BO-led and Cooperative (Natural Language) conditions ($p.\textrm{adj}=0.002$), but not between the Designer-led and Cooperative (Natural Language) conditions ($p.\textrm{adj}=1.000$). Although Q3 was borderline significant ($p=0.050$), none of the pairwise comparisons were significant (Designer-led vs. BO-led: $p.\textrm{adj}=0.659$; Designer-led vs. Cooperative (Natural Language): $p.\textrm{adj}=0.552$; BO-led vs. Cooperative (Natural Language): $p.\textrm{adj}=0.137$), except for a slight tendency between BO-led and Cooperative (Natural Language).

\subsubsection{Participants' Experience in Understanding and Exploring the Design Space}
We asked participants how well they understood the design space and how they explored it. The question items, which are identical to those in \cite{Mo:TIIS:24}, are shown in \autoref{fig:user_study_first_design_space_understanding_and_exploration_experience}, along with boxplots of participants' 7-point Likert ratings across the three conditions (adapted from \cite{Mo:TIIS:24}).

\subsection{Additional Metrics of Pareto Set Discovery}
\label{sec:appendix_additional_metrics_from_user_study_1}

\paragraph{Formal Evaluation}
\figappendixuserstudyfirsttaskperformance

\autoref{fig:appendix_user_study_first_task_performance} (a) shows a boxplot of the total number of formal evaluations performed by each participant in each condition. A repeated measures ANOVA reveals a significant effect of condition ($p=0.000$). Multiple comparisons using paired t-tests with Bonferroni correction indicate significant differences between the Designer-led and BO-led conditions ($p.\textrm{adj}=0.004$) as well as between the BO-led and Cooperative (Natural Language) conditions ($p.\textrm{adj}=0.000$). No significant difference is observed between the Designer-led and Cooperative (Natural Language) conditions ($p.\textrm{adj}=1$).

\paragraph{Pareto Set Count}
\autoref{fig:appendix_user_study_first_task_performance} (b) shows a boxplot of the total number of Pareto-optimal designs obtained. A Friedman test reveals a significant effect of condition ($p=0.043$). Multiple comparisons with Bonferroni correction indicate significant differences between the Designer-led and BO-led conditions ($p.\text{adj}=0.040$). No significant differences are observed between the Designer-led and Cooperative (Natural Language) conditions ($p.\text{adj}=0.776, r=0.257$) or between the BO-led and Cooperative (Natural Language) conditions ($p.\text{adj}=0.133, r=0.465$).

\paragraph{Design Space Count Metric}
\autoref{fig:appendix_user_study_first_task_performance} (c) shows a boxplot of the design space count metric. This metric measures the extent of exploration by dividing the design space into $m$ equal parts for each dimension $d$, creating $m^d$ hypercubes (in our case $m=3$, $d=5$, hence $3^5=243$ hypercubes). The metric counts the number of hypercubes that contain evaluated parameters, using only parameters evaluated through formal evaluation. A larger value indicates a wider exploration. A Friedman test reveals a significant effect of condition ($p=0.0258$). No significant differences are observed between any conditions (Designer-led vs. BO-led: $p.\text{adj}=1.0000$; Designer-led vs. Cooperative (Natural Language): $p.\text{adj}=0.3561$; BO-led vs. Cooperative (Natural Language): $p.\text{adj}=0.0592$).

\paragraph{Total and Mean Design Space Travel Distance}
Two metrics were calculated as proxy measures for assessing the extent of broad exploration versus local fixation. The total travel distance (\autoref{fig:appendix_user_study_first_task_performance} (d)) is the sum of the Euclidean distances between consecutive formally evaluated points in the design space. The mean travel distance (\autoref{fig:appendix_user_study_first_task_performance} (e)) is the travel distance divided by the total number of evaluations performed. A Friedman test reveals a significant effect of condition for total travel distance ($p=0.0000$) and for mean travel distance ($p=0.0003$). Multiple comparisons with Bonferroni correction indicate significant differences for total travel distance between the Designer-led and BO-led conditions ($p.\text{adj}=0.0000$) and between the BO-led and Cooperative (Natural Language) conditions ($p.\text{adj}=0.0002$). No significant difference is observed between the Designer-led and Cooperative (Natural Language) conditions ($p.\text{adj}=0.4621$). For mean travel distance, significant differences are observed between the Designer-led and BO-led conditions ($p.\text{adj}=0.0003$) and between the BO-led and Cooperative (Natural Language) conditions ($p.\text{adj}=0.0058$), while no significant difference is found between the Designer-led and Cooperative (Natural Language) conditions ($p.\text{adj}=0.1027$).

\section{Questionnaire and Detailed Results from User Study 2}

\subsection{Questionnaire}

\subsubsection{Multidimensional Trust Questionnaire}
\figappendixuserstudysecondmtqitems
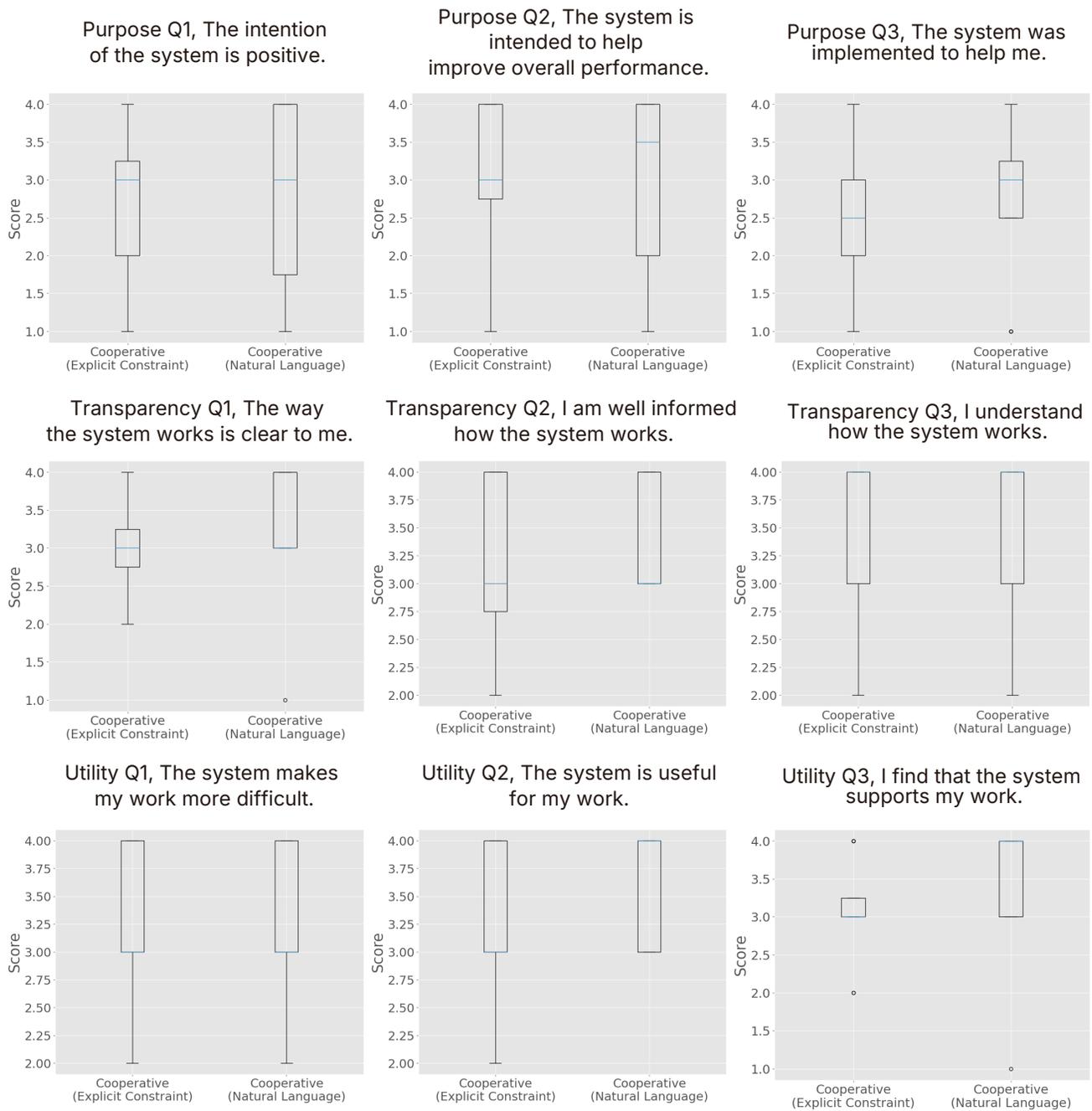
\autoref{fig:appendix_user_study_second_mtq_items} shows the boxplots for participants' 4-point Likert ratings for three subdimensions—purpose, transparency, and utility—of the Multidimensional Trust Questionnaire (MTQ) \cite{Roesler:HFESAM2022}, which was used to measure trust in the system. 

\subsubsection{Agency-Related Questionnaire}
\label{sec:appendix_user_study_2_agency_related_questionnaire}
\figappendixuserstudysecondagency
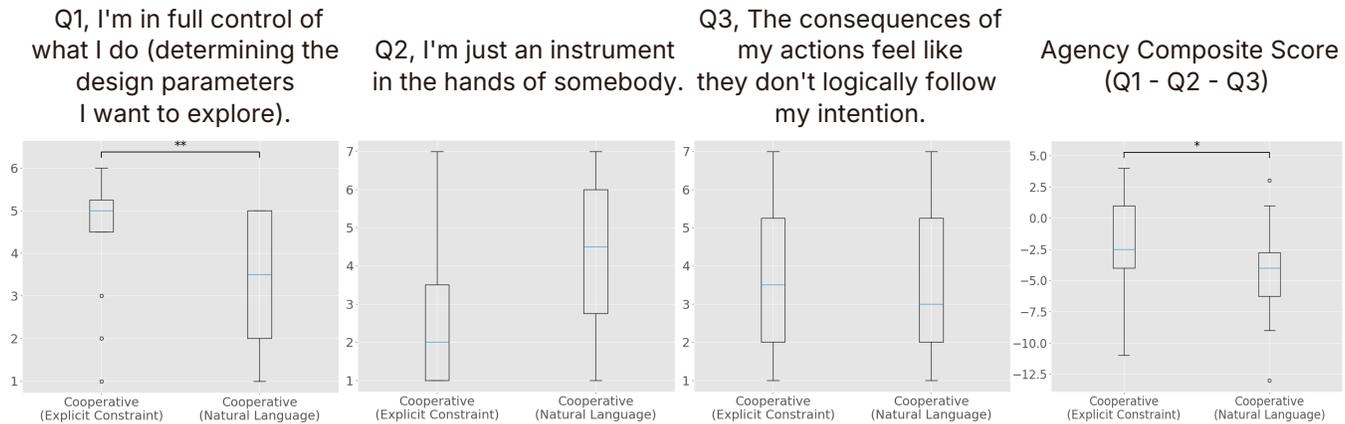
\autoref{fig:appendix_user_study_second_agency_all} shows the boxplots for the three agency-related questions (Q1, Q2, Q3) and the composite score ($Q1 - Q2 - Q3$) in User Study 2. The same items were used in User Study 1.
We found significant differences in Q1 ($p=0.010$; Wilcoxon) and in the overall agency score ($p=0.032$; paired t-test), both favoring the Cooperative-EC (Mo~\etal) condition over the Cooperative-NL condition. In contrast, Q2 ($p=0.061$; Wilcoxon) and Q3 ($p=0.699$; paired t-test) showed no significant differences.

\subsubsection{Participants' Experience in Understanding and Exploring the Design Space}
\figappendixuserstudyseconddesignspaceunderstandingboxplot
As in User Study 1, we asked participants how well they understood the design space and how they explored it. The question items, which are identical to those in \cite{Mo:TIIS:24}, are shown in \autoref{fig:user_study_second_design_space_understanding_and_exploration_experience}, along with the boxplots of participants' 7-point Likert ratings across the three conditions (adapted from \cite{Mo:TIIS:24}).

\subsection{Additional Metrics of Pareto Set Discovery}

Below, we report the same metrics as in the Appendix of User Study 1 (see \label{sec:appendix_additional_metrics_from_user_study_1} for detailed definitions and motivations) and present the statistical outcomes for the two conditions (Cooperative (Explicit Constraints) vs. Cooperative (Natural Language)).

\figappendixuserstudysecondtaskperformance

\paragraph{Formal Evaluation (Total Formal Evaluation Count)} \autoref{fig:appendix_user_study_second_task_performance} (a) shows a boxplot of the total number of formal evaluations. A paired t-test reveals a significant difference between the two conditions ($p=0.036$).

\paragraph{Pareto Set Count} \autoref{fig:appendix_user_study_second_task_performance} (b) shows a boxplot of the total number of Pareto-optimal designs. A Wilcoxon signed-rank test indicates a significant difference ($p=0.009$).

\paragraph{Design Space Count Metric} \autoref{fig:appendix_user_study_second_task_performance} (c) presents the design space count metric. A paired t-test finds no significant difference ($p=0.862$). Hence, there is no evidence that either interface produced broader or narrower exploration in terms of this metric.

\paragraph{Total and Mean Design Space Travel Distance} \autoref{fig:appendix_user_study_second_task_performance} (d) and (e) show, respectively, the total and mean travel distances. Wilcoxon signed-rank tests reveal no significant differences in total travel distance ($p=0.424$) or mean travel distance ($p=0.569$). These results suggest that the overall patterns of exploration or fixation, as reflected by travel distance, did not differ systematically between the two conditions.

\end{document}